\documentclass[aps,prb,graphicx,twocolumn]{revtex4-1}

\usepackage[version=3]{mhchem} % Formula subscripts using \ce{}
\usepackage[T1]{fontenc}
\usepackage{siunitx}
\usepackage{graphicx}

\begin{document}
\title{Graphene Squeeze-Film Pressure Sensors}

\author{Robin J. Dolleman}
\author{Dejan Davidovikj}
\author{Santiago J. Cartamil-Bueno}
\author{Herre S.J. van der Zant}
\author{Peter G. Steeneken}
\affiliation{Kavli Institute of Nanoscience, Delft University of Technology, Lorentzweg 1, 2628CJ, Delft, The Netherlands}
\email{R.J.Dolleman@tudelft.nl}

\begin{abstract}
The operating principle of squeeze-film pressure sensors is based on the pressure dependence of a membrane's resonance frequency, caused by the compression of the surrounding gas which changes the resonator stiffness. To realize such sensors, not only strong and flexible membranes are required, but also minimization of the membrane's mass is essential to maximize responsivity. Here, we demonstrate the use of a few-layer graphene membrane as a squeeze-film pressure sensor. A clear pressure dependence of the membrane's resonant frequency is observed, with a frequency shift of 4 MHz between 8 and 1000 mbar. The sensor shows a reproducible response and no hysteresis. The measured responsivity of the device is 9000 Hz/mbar, which is a factor 45 higher than state-of-the-art MEMS-based squeeze-film pressure sensors while using a 25 times smaller membrane area.
\end{abstract}

\maketitle

\begin{figure}
\includegraphics[scale=1]{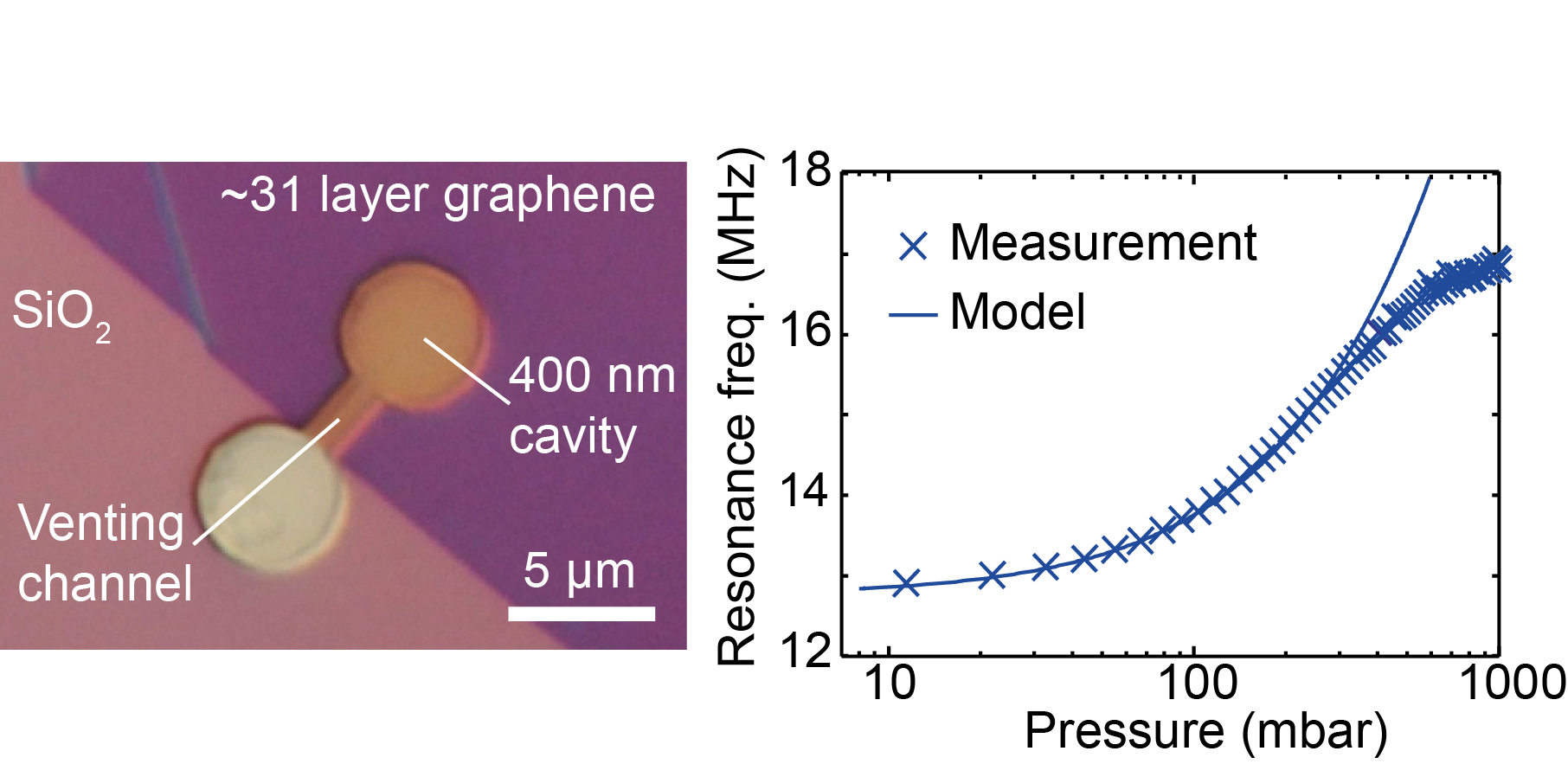}
\end{figure}

Graphene, a single layer of $\mathrm{sp}^2$ bonded carbon atoms\cite{novoselov2005two}, has exceptional mechanical properties. It has the highest Young's modulus ($\sim$1 TPa) of all known materials \cite{lee2008measurement,lee2013high}. Moreover, it has the lowest mass density and bending rigidity of all impermeable membranes\cite{bunch2008impermeable}. These properties make graphene a suitable material for nanomechanical sensors. Currently, pressure sensors are the most widespread membrane-based mechanical sensors and are present in most modern mobile handsets. Commercial microelectromechanical system (MEMS) based pressure sensors feature membranes of several hundreds of nanometers thickness. Replacing these by thin graphene membranes would allow an increase in responsivity and a size reduction by orders of magnitude. In order to exploit these advantages, several studies \cite{bunch2008impermeable,smith2013electromechanical,lee2014atomically,lee2014air} have demonstrated the feasibility of sensing pressure changes with a graphene membrane suspended over a reference cavity at pressure $p_{\mathrm{ref}}$. When the ambient pressure ($p_{\mathrm{amb}}$) changes, the pressure difference ($p_{\mathrm{amb}}-p_{\mathrm{ref}}$) causes a deflection of the membrane. This has been directly detected by atomic force microscopy (AFM) and via a tension induced change in the membrane's resonance frequency \cite{bunch2008impermeable}. Also the change in piezoresistance \cite{smith2013electromechanical} has been used to detect the change in pressure. However, the drawback of these pressure-difference based sensing methods is that they require a stable reference pressure $p_{\mathrm{ref}}$ over the $\sim$10 years lifetime of the sensor, posing extreme demands on the hermeticity of the reference cavity. Even though graphene sealed cavities were shown to have leak time constants of many hours \cite{bunch2008impermeable}, at this stage it is unclear whether these can ever be increased to timescales of years. It is therefore of interest to develop pressure sensors that do not rely on the presence of an impermeable reference cavity.

\begin{figure}
\includegraphics[scale=1]{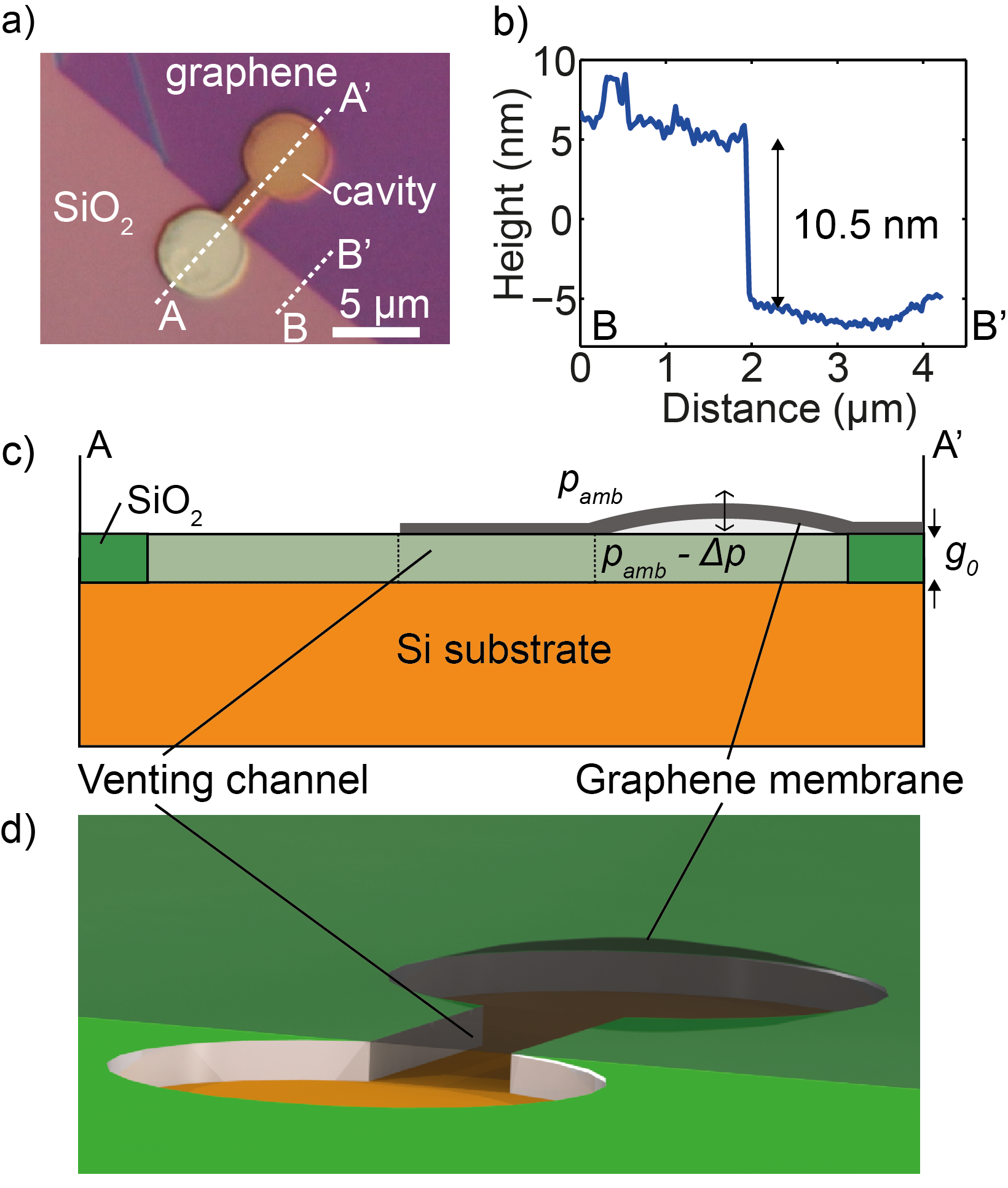}
\caption{a) Graphene flake transferred on a dumbbell shaped hole in a  $\mathrm{SiO}_2$ substrate. Half of the dumbbell is covered, thus creating a drum with a venting channel. The diameter of the drum is 5 $\mu$m and the thickness of the oxide 400 nm. Lines A-A' and B-B' correspond to those in figures c) and b) respectively. b)  Height profile from the atomic force microscopy (AFM) measurement, showing that the membrane is 10.5 $\pm$ 0.7 nm thick. c) Schematic cross-section of the squeeze-film sensor. d) Three-dimensional representation of the squeeze-film sensor design.} \label{fig:fabrication}
\end{figure}

In this work we demonstrate the feasibility of using graphene as a squeeze-film pressure sensor. The sensor consists of a membrane that covers a gas cavity, as is shown in Figure \ref{fig:fabrication}. The main difference with conventional pressure sensors is the presence of an open venting channel that maintains the average pressure inside the cavity equal to the ambient pressure. Squeeze-film pressure sensors operate by compressing gas in the cavity that is at ambient pressure $p_{\mathrm{amb}}$. When the compression is performed at a high frequency, the gas fails to escape its effective position because of the viscous forces \cite{bao2007squeeze}. The added stiffness due to the compression of the gas is a function of pressure. For isothermal compression, this will change the resonance frequency ($f_{\mathrm{res}}$) of the resonator according to:
\begin{equation} \label{eq:squeezefilm}
f_{\mathrm{res}}^2 = f_0^2 + \frac{p_{\mathrm{amb}}}{4\pi^2 g_0 \rho h}.
\end{equation}
Here, $f_{\mathrm{res}}$ is the membrane's resonance frequency at pressure $p_{\mathrm{amb}}$, $f_0$ the resonance frequency in vacuum, $g_0$ the gap size between the membrane and the substrate that lies underneath the membrane and $\rho h$ the mass per unit square (see Supporting Information). Note, that the smaller the  mass per unit square $\rho h$, the larger the frequency shift. The low mass density of graphene thus makes it a perfect material for this type of sensor.

As is shown in the Supporting Information, at high enough frequencies equation \ref{eq:squeezefilm} is independent of mode-shape, thickness and boundary conditions of the membrane. The independence of the boundary conditions shows that the venting channel has no influence on the responsivity ($R={\rm d}f_{\mathrm{res}}/{{\rm d }p_{\mathrm{amb}}}$) of the device. Several works have demonstrated MEMS based squeeze-film pressure sensors with responsivities of up to 200 Hz/mbar \cite{andrews1993resonant,southworth2009pressure,andrews1993comparison,kumar2015mems}. %southworth, figure 1b

We use an exfoliated few-layer graphene (FLG) flake that is suspended over dumbbell shaped holes using a dry stamping method \cite{castellanos2011atomically,castellanos2013single}. The dumbbells have a diameter of 5 $\mu$m and are etched into a 400 nm $\mathrm{SiO}_2$ layer on a silicon substrate (Figure \ref{fig:fabrication}a). The thickness of the flake after transfer is measured to be about 10.5 nm using atomic force microscopy (Figure \ref{fig:fabrication}b). The stamping method allows accurate placement of the flake such that it covers half of the dumbbell shape, thus creating a graphene-based squeeze-film pressure sensor with a lateral venting channel (Figure \ref{fig:fabrication}c--d). To demonstrate the importance of the venting channel for the sensor response, several sealed drums are created with the same flake. The resonance frequency of these sealed drums shows a strikingly different pressure dependence and undesired hysteresis as is discussed in the Supporting Information. 

\begin{figure}
\includegraphics[scale=1]{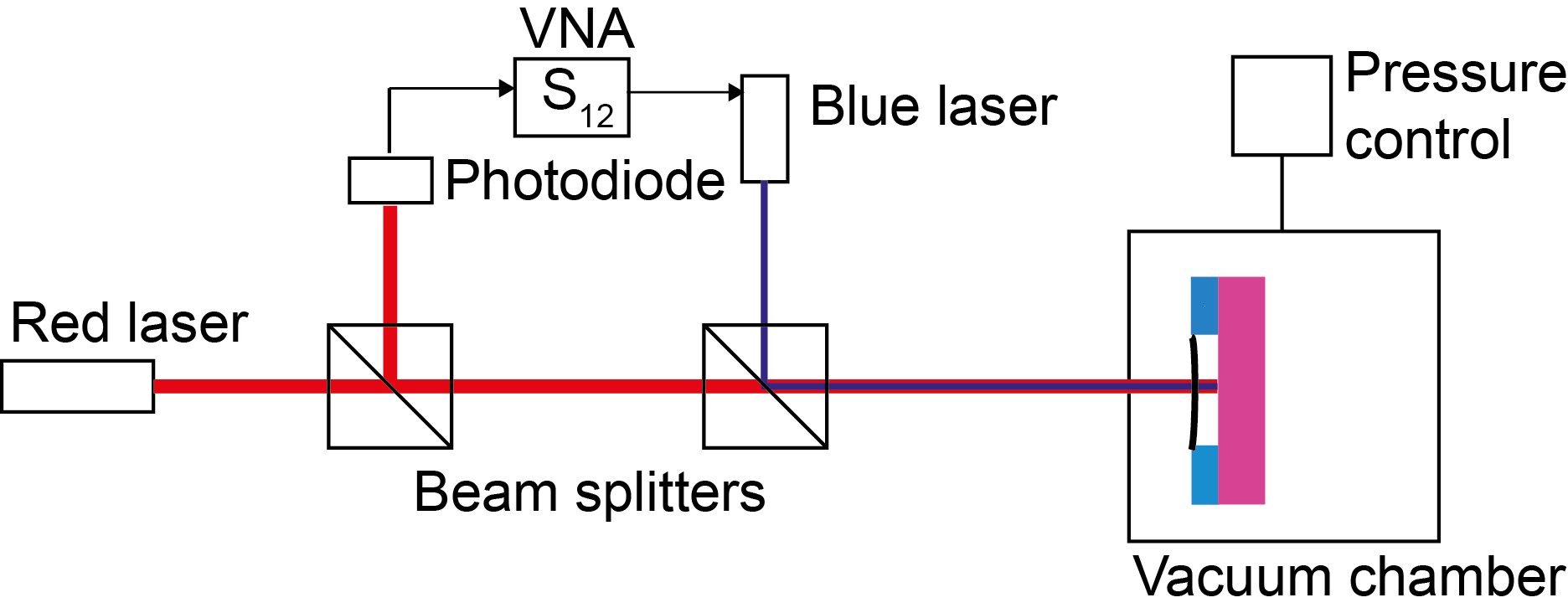}
\caption{Interferometry setup for detecting the resonance frequency of the graphene drum.} \label{fig:setup}
\end{figure}

Figure \ref{fig:setup} shows the interferometry setup that is used to detect the sensor's mechanical resonance modes as a function of gas pressure. The samples are mounted in a vacuum chamber with optical access. A dual-valve pressure controller connected to a nitrogen bottle controls the N$_{\rm 2}$ pressure in the chamber between 8 and 1000 mbar. An intensity modulated 405 nm blue laser drives the graphene membrane via optical absorption and thermomechanical force. A 632 nm red He-Ne laser beam targets the drum and cavity bottom and interference is detected at a photodiode. The intensity is modulated by the mechanical motion of the graphene drum. A vector network analyzer (VNA) modulates the blue diode intensity and detects the red laser light intensity on the photodiode to determine the frequency spectrum of the membrane \cite{castellanos2013single,cartamil2015high}.

\begin{figure}
\includegraphics[scale=1]{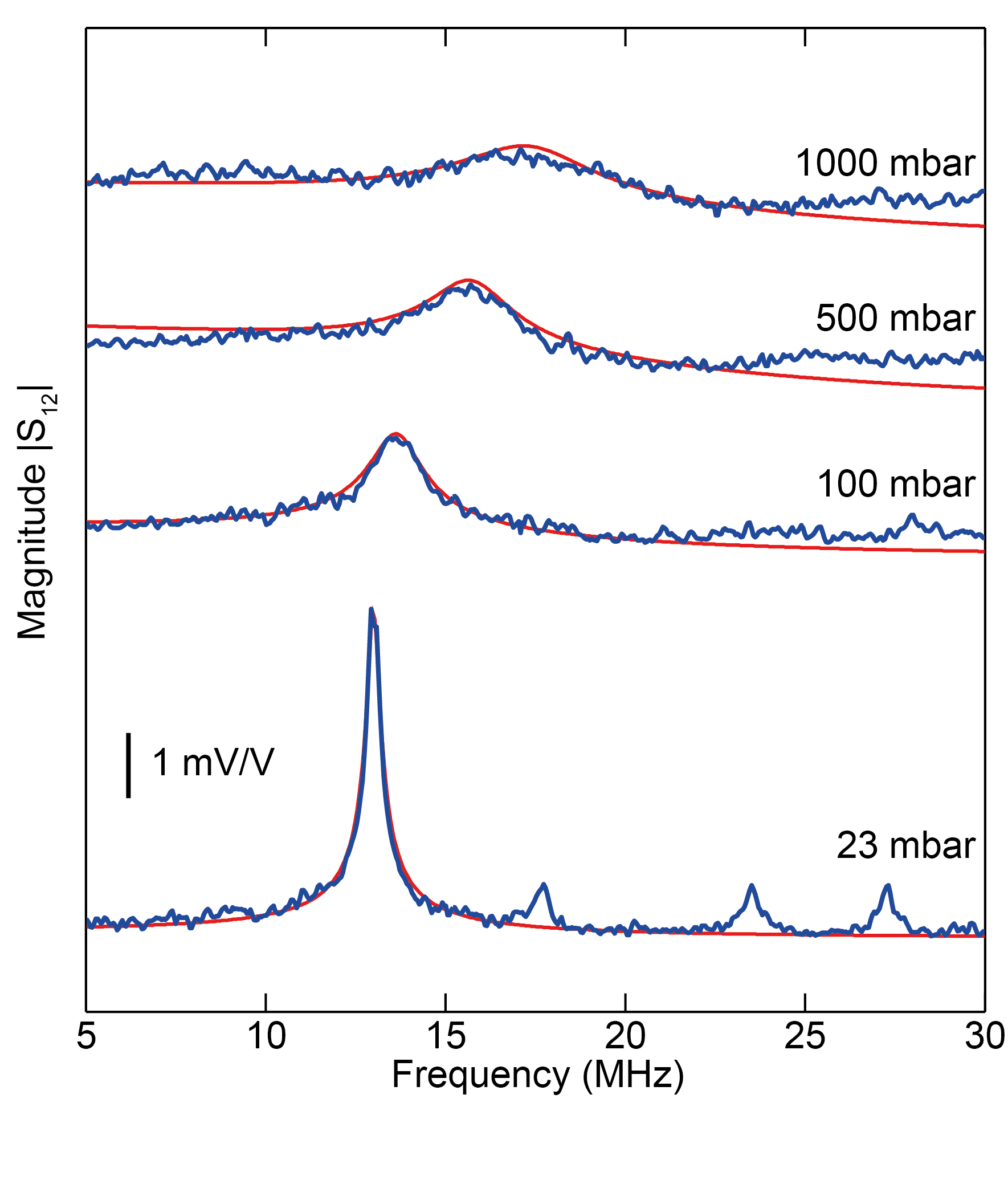}
\caption{\label{fig:datafits} Frequency spectra (blue) obtained from the VNA at different pressures. A damped harmonic oscillator model is fitted (red) to the fundamental mode to determine its resonance frequency and Q-factor.}
\end{figure}

The pressure-dependent resonance frequency of the sensor is studied by ramping the pressure upward and downward at a constant rate. During the pressure ramp the VNA continuously measures frequency spectra from 5-30 MHz at a rate of about 1 sweep every 2 seconds. 
Figure \ref{fig:datafits} shows these frequency spectra at 4 different pressures. At 8 mbar 4 resonance modes are visible. At higher pressures the frequency of the fundamental mode increases while its Q-factor decreases. A damped harmonic oscillator model is fitted (red lines) to the data to extract the resonance frequency and quality factor as a function of pressure. The total frequency shift between 8 mbar and 1000 mbar is 4 MHz.

\begin{figure}
\includegraphics[scale = 1]{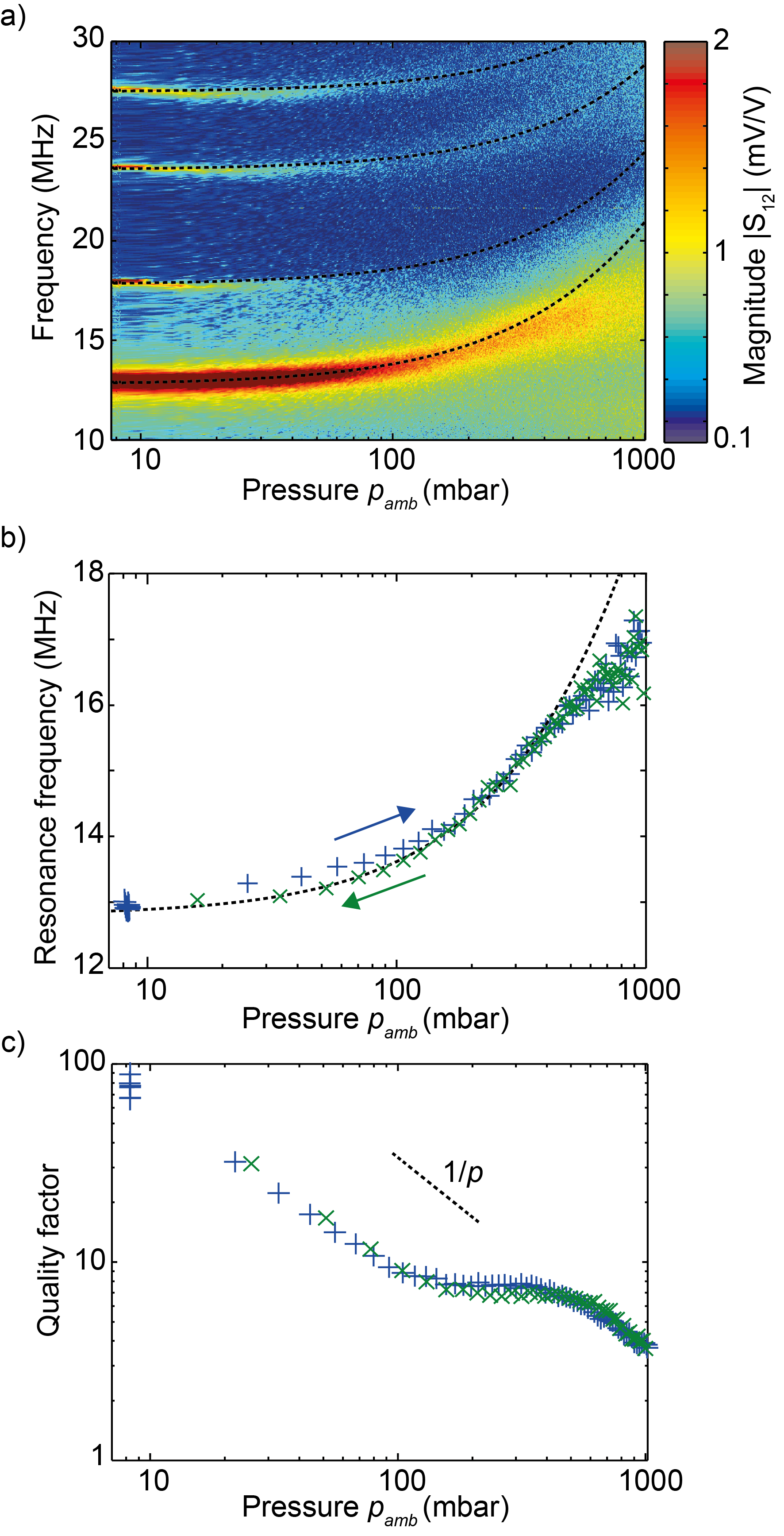}
\caption{Pressure dependent resonances. a) Contour graph of the VNA frequency spectra versus pressure at a ramp rate of 0.55 mbar/s. Dashed black lines are plotted using equation \ref{eq:squeezefilm} with $g_0 = 400~\si{\nano \metre}$ and $\rho h = \num{23.8e-6} ~\si{\kilogram \per \square \metre}$ (31 layers of graphene) using the measured $f_0$ and no fitting parameters. b) Resonance frequencies extracted from an up ($+$) and down ($\times$) pressure sweep at a rate of 3.3 mbar/s, showing the reproducibility of the frequency response. c) Quality factors from an up ($+$) and down ($\times$)  pressure sweep at a rate of 0.55 mbar/s. To reduce the amount of data, the mean of the pressure and average of the quality factor for 10 data points was taken.\label{fig:data}}
\end{figure}

Figure \ref{fig:data}a shows the frequency spectra taken during a pressure ramp in a contour plot. The frequencies of the first, third and fourth resonance modes increase as a function of pressure in close agreement (black dashed lines) with equation \ref{eq:squeezefilm}. The intensity of the second mode vanishes above $\sim$50 mbar; therefore it is not possible to compare its response to equation \ref{eq:squeezefilm}. For all modes the intensity decreases rapidly with pressure. The resonance frequency is plotted versus pressure in Figure \ref{fig:data}b for a measurement at a ramp rate of 3.3 mbar/s. This measurement demonstrates the reproducibility of the sensor, showing no hysteresis as the pressure readings during upward and downward sweep are equal within the inaccuracy of the measurement. Equation \ref{eq:squeezefilm} is plotted (dashed black line) in Figure \ref{fig:data}b using the measured $f_0$ and no additional fit parameters. The theoretical curve is in close agreement with experimental data up to pressures of 200 mbar. We have also measured the pressure response with different gases as shown in the Supporting Information, this shows that the compression is isothermal and the use of equation \ref{eq:squeezefilm} is valid. This also demonstrates that the pressure sensor operates independent of the type of gas. Above 200 mbar, the measured resonance frequency deviates from from equation \ref{eq:squeezefilm}. This indicates that the assumptions underlying this equation cannot account anymore for the resonance frequency behavior at these higher pressures. 

%NOTE: I found a factor 10 error in your expression for the mass density of graphene. Luckily the graphs are ok.

The quality factor is determined from harmonic oscillator fits (Figure \ref{fig:datafits}) and plotted in Figure \ref{fig:data}c. Three regimes can be distinguished: at pressures lower than 100 mbar, the quality factor drops as a function of pressure, approximately proportional to $1/p_{\mathrm{amb}}$. It is predicted by Bao et. al. \cite{bao2002energy} that the quality factor scales with $1/p_{\mathrm{amb}}$ in the free molecular flow regime. % as expected for rarefied gases \cite{stifter2012pressure}.
Between 100 and 500 mbar the quality factor appears to be more or less constant. Above 500 mbar the Q-factor reduces further approximately proportional to $1/p_{\mathrm{amb}}$. More sophisticated modeling is needed to explain the behavior of quality factor as a function of pressure.

% http://iopscience.iop.org/article/10.1088/1742-6596/362/1/012033/pdf;jsessionid=3A3E7307C7CCDCBEC200495B0B1EE9A6.c1

%\begin{figure}
%\includegraphics[scale = 1]{mueffandsigma}
%\caption{a) Effective viscosity as function of chamber pressure. b) Squeeze number as function of chamber pressure, using measured values from the 1000 mbar/30 mins sweep.  \label{fig:sqznumber}}
%\end{figure}

From the data in Figure \ref{fig:data}b it is possible to estimate the responsivity of the device: at low pressures the responsivity is approximately 9000 Hz/mbar while at atmospheric pressure it is 1000 Hz/mbar. The highest reported responsivity in squeeze-film MEMS pressure sensors is $200$Hz/mbar\cite{southworth2009pressure}. The responsivity of the graphene-based sensor is thus a factor of $\sim$5-45 larger than that of a MEMS sensor. At the same time the area of the graphene sensor is a factor 25 smaller. 
Based on equation \ref{eq:squeezefilm}, further improvement of the demonstrated sensor concept is possible by reducing the thickness of the membranes. It is estimated that using a single-layer graphene resonator will increase the responsivity by a factor of 5.6. A reduction of the gap size $g_0$ can enable a further increase of the responsivity.

In summary, a graphene-based squeeze-film pressure sensor has been demonstrated that does not need an impermeable reference cavity at stable reference pressure. Reproducible sensor response is demonstrated and a 4 MHz resonance frequency shift between 8 and 1000 mbar is measured. The resonance frequency closely follows the squeeze-film model up to 200 mbar, but at higher pressures deviations from the model are observed that require further theoretical study. In comparison with MEMS based squeeze-film sensors, the responsivity of the sensor is a factor 5-45 larger at an area of a factor 25 smaller. A further increase of the responsivity can be obtained using thinner membranes and reducing the gap size. In comparison to other graphene-based pressure sensing concept, the squeeze-film pressure sensor has the advantage that it does not rely on an impermeable reference cavity at constant pressure. It therefore provides a promising route towards size reduction and sensitivity improvements of pressure sensors.

\section*{Acknowledgements}
We thank Allard Katan, Ronald van Leeuwen and Warner J. Venstra for experimental support. The authors further thank the Dutch Technology Foundation (STW), which is part of the Netherlands Organisation for Scientific Research (NWO), and which is partly funded by the Ministry of Economic Affairs, for financially supporting this work. The research leading to these results has also received funding from the European Union Seventh Framework Programme under grant agreement no 604391 Graphene Flagship and this work was supported by the Netherlands Organisation for Scientific Research (NWO/OCW), as part of the Frontiers of Nanoscience program.

\section*{Methods}
Substrates were fabricated using p-type silicon with thermally grown silicon dioxide on top. 400 nm cavities were etched into the silicon dioxide using reactive ion etching to obtain vertical etch profiles, using chromium as a mask and gases Ar at 2.7 sccm and $\mathrm{CHF}_3$ at 50 sccm with a power of 50 W and 7 $\mathrm{\mu}$bar pressure. The chromium mask was removed and samples were cleaned in nitric acid. For details on the graphene transfer process and the laser interferometer setup the reader is referred to Castellanos-Gomez et. al. \cite{castellanos2013single}. In the measurements the red laser power was 3 mW as measured before the objective entrance. The blue laser was kept at a power less than 1  mW, which was modulated using the VNA at -10 dB, resulting in a power modulation of 4.5\%.  Atomic force microscopy was performed on a Bruker Multimode 3 system to measure the thickness of the graphene flake. Pressure was controlled using a dual-valve pressure controller, calibrated for a pressure range between 0 and 1 bar. Dry nitrogen, carbon dioxide or argon was offered at the IN port of the controller and a (dry) scroll pump was connected to the vacuum port. 
%\appendix
\onecolumngrid
\pagebreak

\setcounter{equation}{0}
\setcounter{figure}{0}
\setcounter{table}{0}
\makeatletter
\renewcommand{\theequation}{S\arabic{equation}}
\renewcommand{\thefigure}{S\arabic{figure}}
\section*{Supporting Information: Graphene Squeeze-Film Pressure Sensors}

\section{Optical image of the flake}
\begin{figure}[b]
\includegraphics[width = 0.6\linewidth]{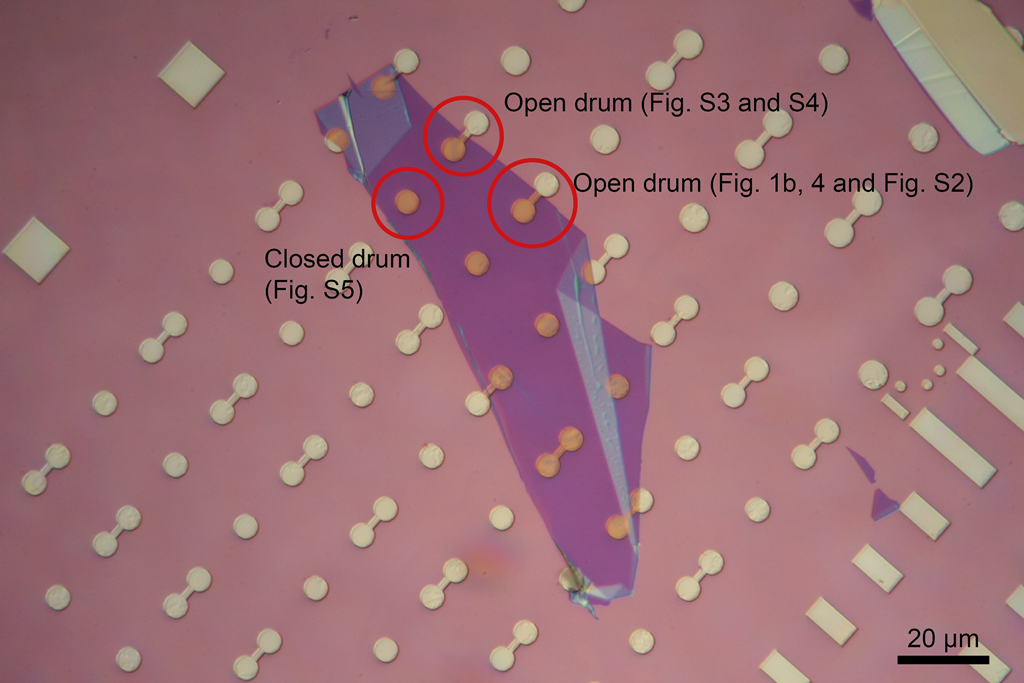}
\caption{Optical image of the flake used in the experiment, highlighted are the drums used in the article (Figure 1b and \ref{fig:figure4add}), another open drum (Figure \ref{fig:otherdrum}) and the closed drum used for reference measurements (Figure \ref{fig:otherdrum}).\label{fig:flake-picture}}
\end{figure}

In Figure \ref{fig:flake-picture} an optical image of the flake used in this work is shown. Both closed drums and drums with venting channels are created with the same flake, ensuring that the mass per unit square is equal for each membrane. Measurement results are presented of the drums that are encircled in the figure.

\section{Additional measurement results}
In this section additional measurements are presented. All measurements were performed on the same graphene flake shown in Figure \ref{fig:flake-picture}. The quality factors corresponding to Figure 4b and frequencies corresponding to Figure 4c in the main text are shown in Figure \ref{fig:figure4add}. The measurement are reproduced on a different open drum as shown in Figure \ref{fig:otherdrum}. Also we measured the response of an open drum with different gases, showing that the compression is isothermal (Figure \ref{fig:isothermal}). A closed drum was used as a reference to the measurement on the open drum. The response of the closed drum is very similar to the one reported by Bunch et. al. \cite{bunch2008impermeable}, with high frequency shifts, as shown in Figure \ref{fig:closeddrum}. However a large hysteresis is found due to leakage, making use as a pressure sensor impossible.

\subsection{Additional measurement results corresponding to Figure 4}
\begin{figure}
\includegraphics[scale = 1]{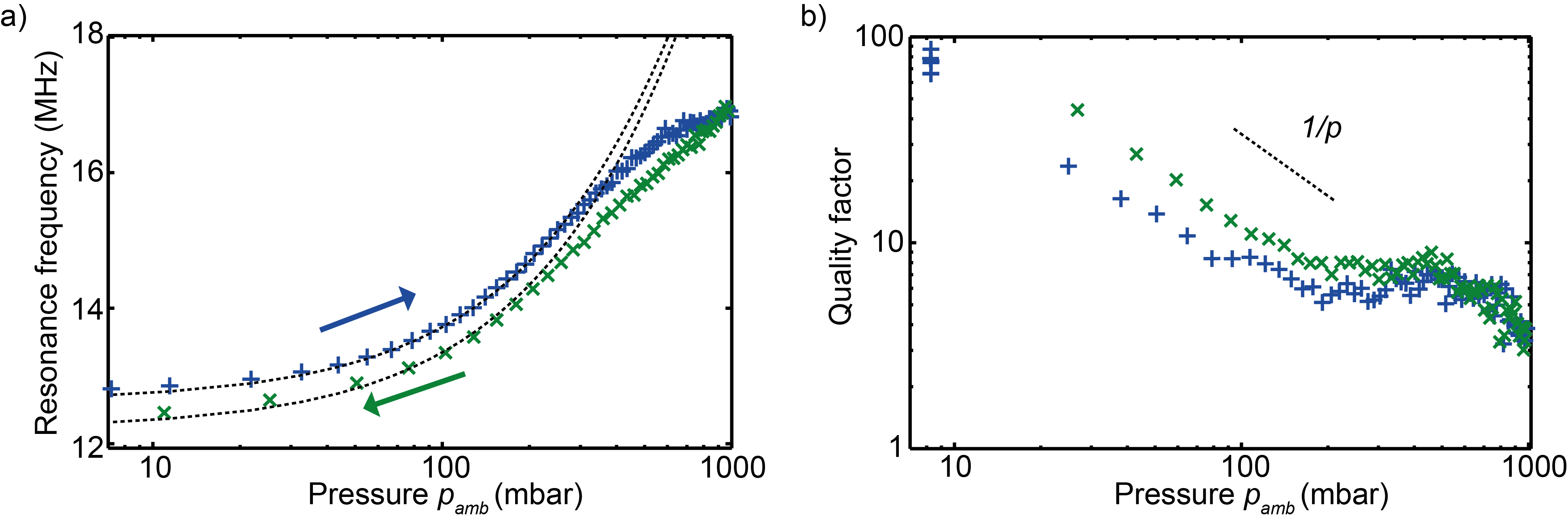}
\caption{a) Frequency response of a measurement taken with a sweep rate of 0.55 mbar/s; the quality factors from this measurement are shown in Figure 4c in the main text. b) Quality factors from a measurement taken with a different ramp rate of 3.3 mbar/s, the frequencies are shown in Figure 4b in the article. \label{fig:figure4add}}
\end{figure}

Additional graphs are presented corresponding to the measurement data shown in Figure 4 of the main text. Figure \ref{fig:figure4add}a shows the frequencies corresponding to the quality factors in Figure 4c, with the pressure ramping at 0.55 mbar/s. Drift is observed, which we attribute to movements in the measurement setup, which change the position of the laser spot on the membrane in the course of the one hour measurement. This modifies the way the substrate thermally expands, thereby changing the tension in the membrane. The drift observed in Figure \ref{fig:figure4add}a is therefore an expected inaccuracy of the measurement. The data fits (dashed lines) are produced using equation 1 from the main text with $f_0 = 12.3$ MHz and $f_0 = 12.7$ MHz to correct for the drift.

Figure \ref{fig:figure4add}b shows the quality factors corresponding to Figure 4b in the main text. The pressure was ramped at a rate of 3.3 mbar/s. The data is in good agreement with Figure 4c in the main text.

\subsection{Measurement on different open drum}
\begin{figure}[h!]
\includegraphics[scale = 1]{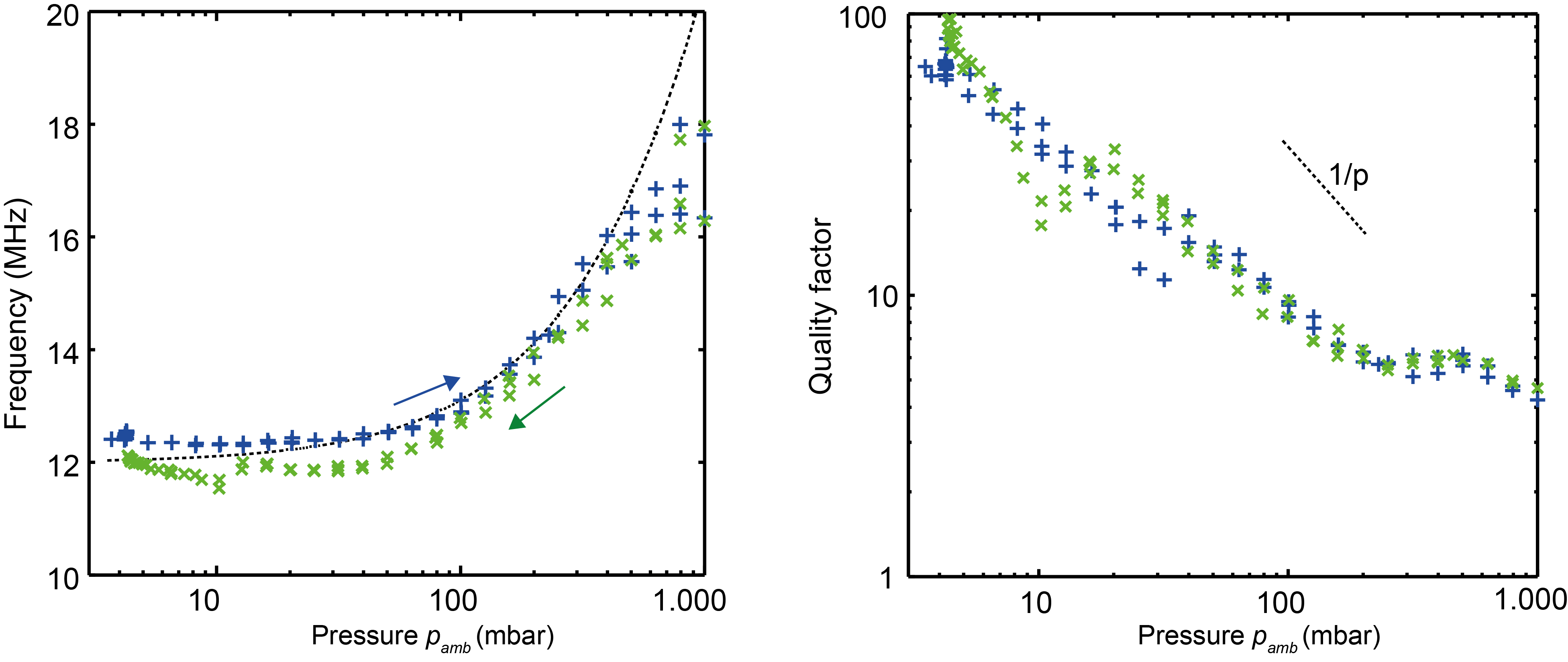}
\caption{Frequency response and quality factor for another open drum on the same flake (see Figure \ref{fig:flake-picture}). \label{fig:otherdrum}}
\end{figure}

In this section measurement results are presented on a different open drum than the one used in the main text (see Figure \ref{fig:flake-picture}). It is found that the response is very similar as is shown in Figure \ref{fig:otherdrum}.  The pressure was changed in logarithmic steps between 3 and 1000 mbar, both upwards and downwards. The total duration of the measurement was 900 seconds. The frequency response is very similar to the open drum used in the main text, within the inaccuracy of the measurement. The quality factor shows a slightly different slope than the other open drum.

\subsection{Measurements with different gases}
\begin{figure}
\includegraphics[scale=1]{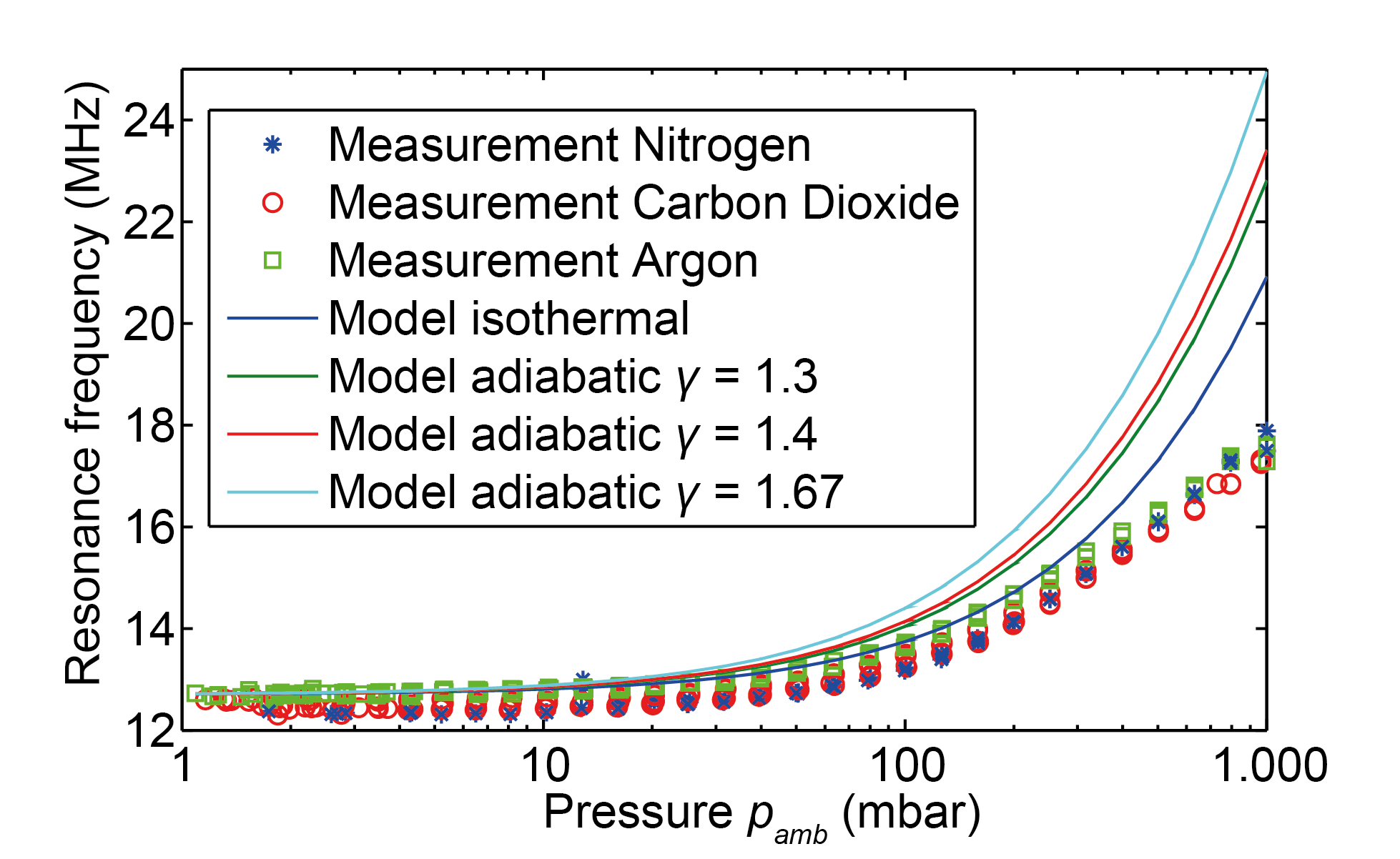}
\caption{Frequency response for nitrogen, argon and carbon dioxide in the chamber. Solid lines are models for isothermal compression ($\gamma = 1$) and adiabatic compression for argon ($\gamma = 1.3$), nitrogen ($\gamma = 1.4$) and carbon dioxide ($\gamma =1.67$). \label{fig:isothermal}} 
\end{figure}

\begin{figure}[h!]
\includegraphics[width = 0.4\linewidth]{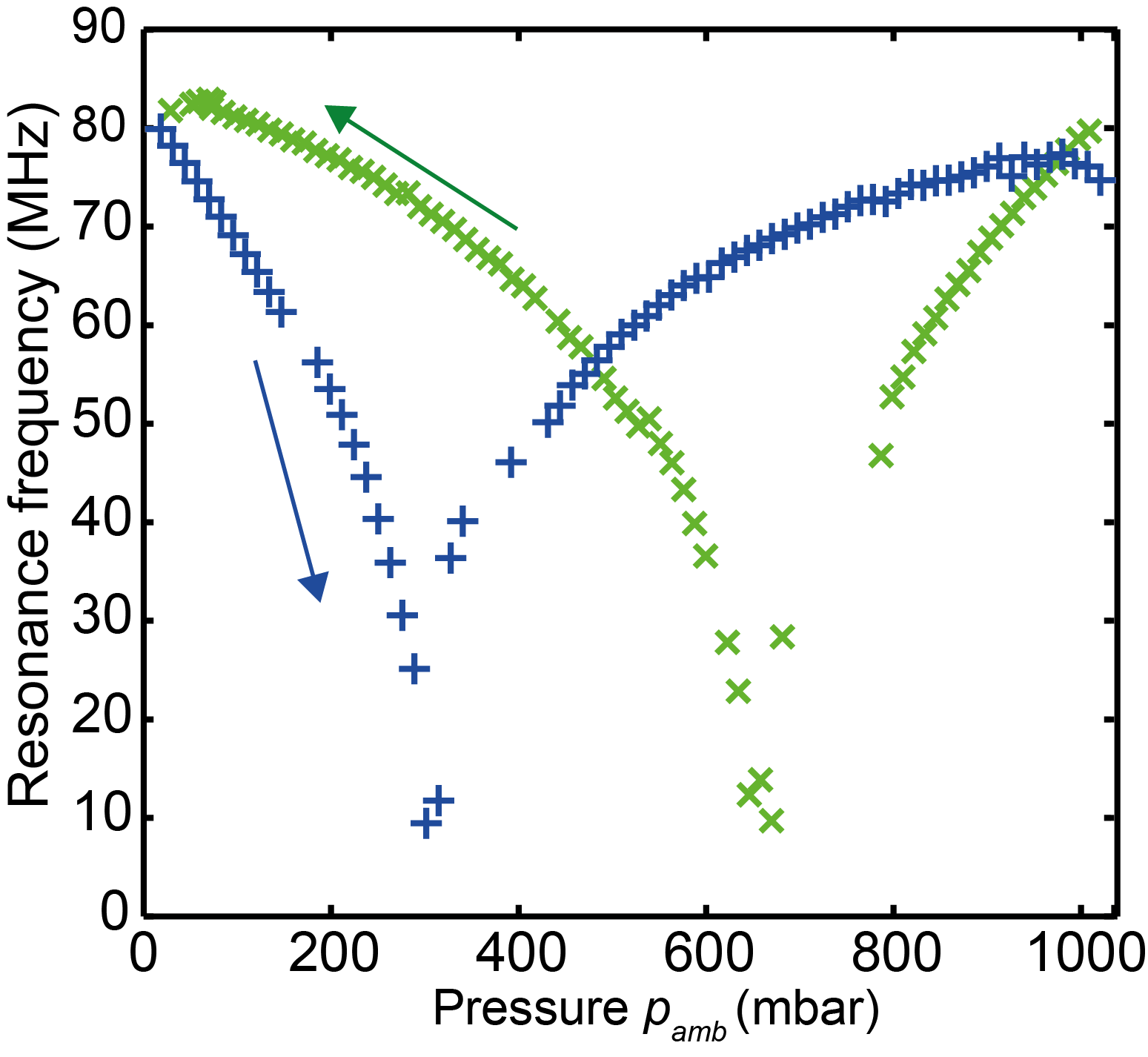}
\caption{Frequency response of a closed drum as a function of pressure. \label{fig:closeddrum}}
\end{figure}

In this section measurement results are presented with different gases, these measurements willl show whether compression in these systems is isothermal or adiabatic. According to Andrews et al. \cite{andrews1993comparison} the response of frequency versus pressure for the case of adiabatic compression is given by:
\begin{equation}
\omega^2 = \omega_0^2 + \gamma \frac{p_\mathrm{amb}}{g_0 \rho h}, 
\label{eq:adiabatic}
\end{equation}
where $\gamma$ is the adiabatic index. For monatomic ideal gases such as argon, $\gamma = 1.3$, for a diatomic gas such as nitrogen, $\gamma = 1.4$ and for a collinear molecule such as carbon dioxide $\gamma = 1.67$. For isothermal processes one can use $\gamma = 1$ independent on the gas used, which makes equation \ref{eq:adiabatic} equal to equation 1 from the main text. In order to examine whether the compression in graphene-based squeeze-film sensors is adiabatic or isothermal, pressure sweeps were performed on the same open drum as Figure \ref{fig:otherdrum} using three gases with different adiabatic indexes (Figure \ref{fig:isothermal}). This measurement shows that compression in graphene squeeze-film pressure sensors is isothermal, since no significant change in stiffness is observed with different gases. This is expected for rarefied gas since the collision frequency is in the order: $f_{\mathrm{col}} = v_{\mathrm{gas}}/g_0 \approx 400/400*10^{-9} = 1$GHz, where  $v_{\mathrm{gas}}$ is the velocity of gas particles. When the gas is not rarefied  $f_{\mathrm{col}} = v_{\mathrm{gas}}/\lambda \approx  400/70*10^{-9} = 5.7$GHz, since this is much larger than the compression frequency, isothermal compression is expected in the cavity over the whole pressure range.

\subsection{Measurement results on a closed drum}
To compare the response of the open drums with those of closed drums, measurements were perfomed on a drum without venting channel on the same flake with equal diameter (see Figure \ref{fig:flake-picture}). The pressure was ramped up and down at a rate of 0.55 mbar/s. The frequency response is strikingly different from the ones observed in open drums, with a clear hysteresis caused by gas leakage of the cavity. As shown in Figure \ref{fig:otherdrum}, the frequency shifts observed are around 85 MHz, much larger than for the squeeze-film effect. It is concluded that these shifts are tension-induced by the pressure difference over the membrane, in agreement with observations by Bunch et. al. \cite{bunch2008impermeable}.

%\begin{figure}[h!]
%\includegraphics[width = 0.6\linewidth]{supplementary-5}
%\caption*{Figure S6: Frequency response of the closed drum, right after the pressure is dropped from 1000 mbar to 8 mbar. The relaxation of the frequency shows that the cavity is leaking, which is causing the hysteresis observed in Figure S4.}
%\end{figure}

\section{Analysis}
In this section equation 1 from the main text is derived using the equations of motion for a piston and a membrane combined with Boyle's law of an infinitesimal part of the gas film. For the derivation of this equation it is assumed that the compression frequency is so high that the gas effectively has no time to allow for significant lateral gas flow within 1 period. The validity of this assumption is investigated using the linearized Reynolds equation in the second part of this section. In the entire analysis only the gas dynamics in the thin gas film underneath the membrane is considered.

\subsection{Derivation of frequency-pressure relation for a piston}
\begin{figure}
\includegraphics{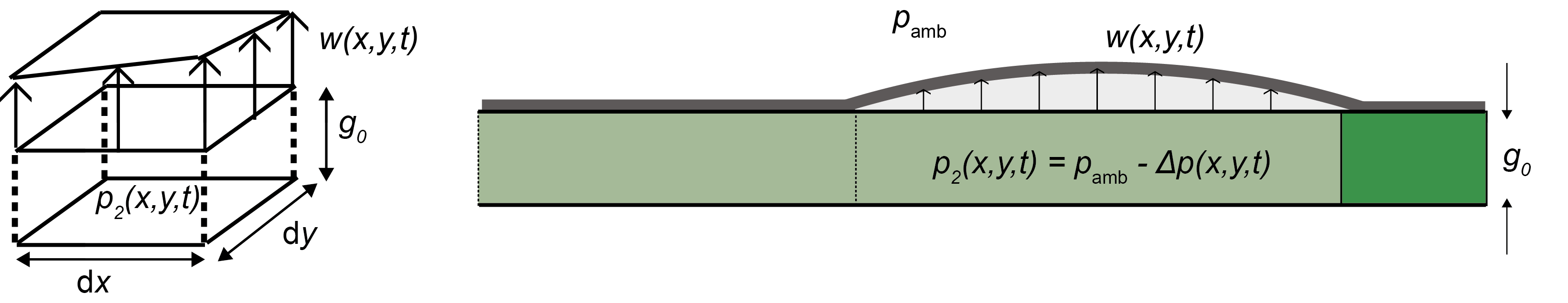}
\caption{Infinitesimal part of the fluid film beneath the resonator and cross section of the sensor. \label{fig:der}}
\end{figure}
In this section equation 1 from the main text is derived, first for a piston followed by the derivation for a membrane.
Assuming ideal compression in the squeeze-film, Boyle's law can be applied to an infinitesimal part of the film (Figure \ref{fig:der}):
\begin{equation}
p_{\mathrm{amb}} V_1 = p_2 V_2,
\end{equation}
\begin{equation}
p_{\mathrm{amb}} g_0 \mathrm{d}x \mathrm{d} y = p_2 (g_0 + w) \mathrm{d} x \mathrm{d} y,
\end{equation}
\begin{equation}
p_{\mathrm{amb}} g_0 = p_2 (g_0 +w),
\end{equation}
by substituting $p_2  =p_{\mathrm{amb}} - \Delta p$:
\begin{equation}
p_{\mathrm{amb}} w = \Delta p g_0 + \Delta p w.
\end{equation}
Assuming $\Delta p <<p_{\mathrm{amb}}$ and $w << g_0$ we obtain the following expression:
\begin{equation} \label{eq:presdefl}
\Delta p = \frac{p_{\mathrm{amb}}}{g_0} w,
\end{equation}
which gives the pressure field, which is in turn proportional to the deflection field. This result is equivalent to the one that was obtained by Bao and Yang for a squeeze-film between two rigid plates\cite{bao2007squeeze}.
The equation of motion for a piston can then be written as\cite{rao2007vibration}:
\begin{eqnarray}\label{eq:eqmotion}
\rho h \frac{\mathrm{d}^2 w}{\mathrm{d} t^2} =  -\Delta p = -\frac{p_{\mathrm{amb}}}{g_0} w.
\end{eqnarray}
For which one can directly obtain the resonance frequency:
\begin{equation}
\frac{\mathrm{d}^2 w}{\mathrm{d} t^2} + \frac{p_{\mathrm{amb}}}{g_0 \rho h} w = 0. 
\end{equation}
This is the equation for a harmonic oscillator with frequency:
\begin{equation}\label{eq:presresp}
\omega^2 =  \frac{p_{\mathrm{amb}}}{g_0 \rho h},
\end{equation}
which gives the pressure response $\Delta \omega = \sqrt{p_{\mathrm{amb}}/g_0 \rho h}$.

\subsubsection{Derivation of the frequency-pressure relation for a flexible membrane}
In this section we derive the frequency pressure relation for a membrane that has both tension (or compression) $n_0$ and bending rigidity $D$. This analysis shows that the response of the frequency is equal to the situation of a piston. The equation of motion is given by \cite{rao2007vibration}:
\begin{eqnarray}
\rho h \frac{\partial^2 w}{\partial t^2} + D \nabla^4 w  - n_0 \nabla^2 w =  -\frac{p_{\mathrm{amb}}}{g_0} w.\label{eq:motionend}
\end{eqnarray}
Equation \ref{eq:motionend} can be solved by separation of variables:
\begin{equation}
w(x,y,t) =W(x,y) T(t),
\end{equation}
\begin{eqnarray}
 \frac{\rho h}{D} \frac{1}{T} \frac{\mathrm{d}^2 T}{\mathrm{d} t^2} + \frac{p_{\mathrm{amb}}}{g_0 D}= \frac{n_0}{D} \frac{\nabla^2 W}{W} - \frac{\nabla^4 W}{W} = \lambda^4, \label{eq:sepend}
\end{eqnarray}
with $\lambda$ as the separation variable. From equation \ref{eq:sepend} one can obtain the time-dependent equations which will be used to calculate the eigenfrequency:
\begin{eqnarray}
\frac{\mathrm{d}^2 T}{\mathrm{d} t^2}+ \left(\frac{p_{\mathrm{amb}}}{g_0 \rho h}  + \lambda^4 \frac{D}{\rho h} \right) T =0. 
\end{eqnarray}
This equation describes an harmonic oscillator:
\begin{equation}
\frac{\mathrm{d}^2 T}{\mathrm{d}^2 t} + \omega^2 T = 0, 
\end{equation}
which means that the resonance frequencies become:
\begin{eqnarray}
\omega^2 = \frac{p_{\mathrm{amb}}}{g_0 \rho h}  +  \lambda^4 \frac{D}{\rho h}.
\end{eqnarray}
If the frequency in vacuum ($p_{\mathrm{amb}}=0$) is written as $\omega_0$, the resonance frequency as function of pressure can be written as:
\begin{equation}\label{eq:squeezefilm}
\omega^2 = \omega_0^2 + \frac{p_{\mathrm{amb}}}{g_0 \rho h}.
\end{equation}
The result is consistent with the one obtained by Andrews\cite{andrews1993resonant} for a square plate and equal to the pressure response of a piston since $\Delta \omega^2 = p_{\mathrm{amb}}/(g_0 \rho h)$. Note, that the shape, boundary conditions, thickness and tension do influence $\omega_0$, but not the pressure response. Only the mass of the membrane and gap size influence the pressure response. This is a useful property, since the behavior of the sensor can be predicted by measuring the gap size, flake thickness and frequency at vacuum, which simplifies the analysis of the measurement.

From equation \ref{eq:sepend} a useful identity can derived that will be used in the next section:
\begin{eqnarray}
\left( \nabla^4 - \frac{n_0}{D}\nabla^2 - \lambda^4 \right) W = 0,\\
\left(\nabla^2 + \alpha^2\right)\left(\nabla^2 - \left(\frac{n_0}{D} +\alpha^2\right)\right) W = 0,\\
\alpha^2= -\frac{\nabla^2 W}{W} \mathrm{~and~} \frac{n_0}{D} + \alpha^2=  \frac{\nabla^2 W}{W}.  \label{eq:lambdaend}
\end{eqnarray}
Note, that as a result of the separation of variables the ratio $\frac{\nabla^2 W}{W}$ becomes a constant that is no longer dependent on position. In equations \ref{eq:sepend} and \ref{eq:lambdaend}, the relation between the constants $\alpha$ and $\lambda$ is:
\begin{equation}
 \lambda^2 = \alpha^2 \sqrt{\frac{n_0}{D} + \alpha^2}.
\end{equation}
For circular plates we can write:
\begin{eqnarray} \label{eq:gamma}
\alpha = \frac{\gamma_{mn}}{a}, \label{eq:gammaend}
\end{eqnarray}
where $\gamma_{mn}$ is the root of the frequency relation \cite{wah1962vibration}, which depends on the boundary conditions and mode-shape of the diaphragm. $a$ is the radius of the diaphragm.

\section{Frequency range for compression in squeeze-film sensors}
In the derivation of the pressure response of the sensor (equation \ref{eq:squeezefilm}) it is assumed that the gas is compressed at very high frequency. In this section we derive the minimal frequency at which this assumption is valid. For this purpose, the linear Reynolds equation is written as\cite{bao2007squeeze}:
\begin{equation}
p_{\mathrm{amb}} \nabla^2 p -\frac{12\mu}{g_0^2} \frac{\partial p}{\partial t} = \frac{12 \mu p_{\mathrm{amb}}}{g_0^3} \frac{\partial g}{\partial t}.
\end{equation}
Use $g = g_0+w$ and substitute equation \ref{eq:presdefl}:
\begin{equation}
\frac{p_{\mathrm{amb}}^2}{g_0} \nabla^2 w + \frac{12 \mu p_{\mathrm{amb}}}{g_0^3} \frac{\partial w}{\partial t} =  \frac{12 \mu p_{\mathrm{amb}}}{g_0^3} \frac{\partial w}{\partial t}. 
\end{equation}
The left and right side become equal when the following condition applies:
\begin{equation}\label{eq:omegabig}
\frac{p_{\mathrm{amb}}^2}{g_0} \nabla^2 w <<\frac{12 \mu p_{\mathrm{amb}}}{g_0^3} \frac{\partial w}{\partial t}.
\end{equation}
Since $w$ is assumed to undergo harmonic motion, one can write $w = C W(x,y)\sin{\omega t}$, where $C$ is the amplitude and $W(x,y)$ the mode-shape. Using $\frac{\nabla^2 W}{W} = - \frac{\gamma_{mn}^2}{a^2}$  from the previous section, we can write for circular diaphragms:
\begin{equation}\label{eq:omegac}
\omega >>\omega_c= \frac{p_{\mathrm{amb}} g_0^2 \gamma_{mn}^2}{12 \mu a^2}. 
\end{equation}
\begin{figure}[h!]
\includegraphics[scale=1]{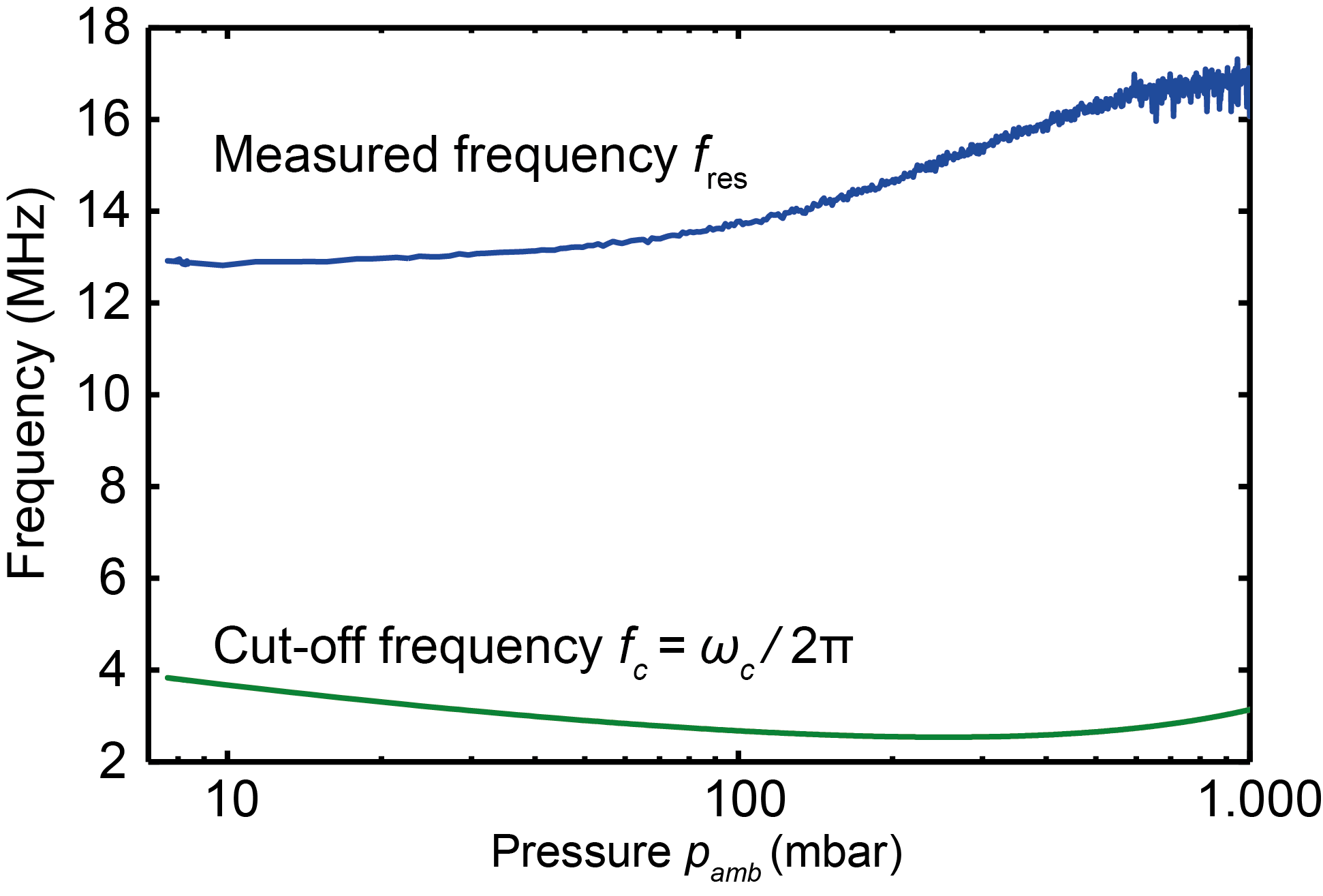}
\caption{Measured resonance frequencies compared to the cut-off frequency (equation \ref{eq:omegac}).\label{fig:omegac}}
\end{figure}
The frequency from Figure \ref{fig:figure4add}a (upwards sweep) is plotted in Figure \ref{fig:omegac} and compared to equation \ref{eq:omegac}. The measured frequency is much higher than the cut-off frequency, which shows that equation \ref{eq:squeezefilm}. Since the mean free path of the gas molecules is of the same order as the dimensions of our device, it is no longer valid to use the bulk viscosity ($\mu_0$). Instead, the model proposed by Veijola\cite{veijola1995equivalent} is used to approximately correct the viscosity ($\mu_{\mathrm{eff}}$) with the formula:
\begin{equation}
\mu_{\mathrm{eff}} = \frac{\mu_0}{1+9.638\mathrm{Kn}^{1.159}},
\end{equation}
where $\mathrm{Kn}$ is the Knudsen number defined as the ratio between the mean free path and the gap size. Note, that this model loses its validity at pressures lower than approximately 30 mbar, since the mean free path becomes of the same order as the diameter of the cavity. More sophisticated modeling is necessary to determine a more appropriate effective viscosity for this situation.

\bibliography{squeezefilm}

\begin{thebibliography}{19}%
\makeatletter
\providecommand \@ifxundefined [1]{%
 \@ifx{#1\undefined}
}%
\providecommand \@ifnum [1]{%
 \ifnum #1\expandafter \@firstoftwo
 \else \expandafter \@secondoftwo
 \fi
}%
\providecommand \@ifx [1]{%
 \ifx #1\expandafter \@firstoftwo
 \else \expandafter \@secondoftwo
 \fi
}%
\providecommand \natexlab [1]{#1}%
\providecommand \enquote  [1]{``#1''}%
\providecommand \bibnamefont  [1]{#1}%
\providecommand \bibfnamefont [1]{#1}%
\providecommand \citenamefont [1]{#1}%
\providecommand \href@noop [0]{\@secondoftwo}%
\providecommand \href [0]{\begingroup \@sanitize@url \@href}%
\providecommand \@href[1]{\@@startlink{#1}\@@href}%
\providecommand \@@href[1]{\endgroup#1\@@endlink}%
\providecommand \@sanitize@url [0]{\catcode `\\12\catcode `\$12\catcode
  `\&12\catcode `\#12\catcode `\^12\catcode `\_12\catcode `\%12\relax}%
\providecommand \@@startlink[1]{}%
\providecommand \@@endlink[0]{}%
\providecommand \url  [0]{\begingroup\@sanitize@url \@url }%
\providecommand \@url [1]{\endgroup\@href {#1}{\urlprefix }}%
\providecommand \urlprefix  [0]{URL }%
\providecommand \Eprint [0]{\href }%
\providecommand \doibase [0]{http://dx.doi.org/}%
\providecommand \selectlanguage [0]{\@gobble}%
\providecommand \bibinfo  [0]{\@secondoftwo}%
\providecommand \bibfield  [0]{\@secondoftwo}%
\providecommand \translation [1]{[#1]}%
\providecommand \BibitemOpen [0]{}%
\providecommand \bibitemStop [0]{}%
\providecommand \bibitemNoStop [0]{.\EOS\space}%
\providecommand \EOS [0]{\spacefactor3000\relax}%
\providecommand \BibitemShut  [1]{\csname bibitem#1\endcsname}%
\let\auto@bib@innerbib\@empty
%</preamble>
\bibitem [{\citenamefont {Novoselov}\ \emph {et~al.}(2005)\citenamefont
  {Novoselov}, \citenamefont {Geim}, \citenamefont {Morozov}, \citenamefont
  {Jiang}, \citenamefont {Katsnelson}, \citenamefont {Grigorieva},
  \citenamefont {Dubonos},\ and\ \citenamefont {Firsov}}]{novoselov2005two}%
  \BibitemOpen
  \bibfield  {author} {\bibinfo {author} {\bibfnamefont {K.}~\bibnamefont
  {Novoselov}}, \bibinfo {author} {\bibfnamefont {A.~K.}\ \bibnamefont {Geim}},
  \bibinfo {author} {\bibfnamefont {S.}~\bibnamefont {Morozov}}, \bibinfo
  {author} {\bibfnamefont {D.}~\bibnamefont {Jiang}}, \bibinfo {author}
  {\bibfnamefont {M.}~\bibnamefont {Katsnelson}}, \bibinfo {author}
  {\bibfnamefont {I.}~\bibnamefont {Grigorieva}}, \bibinfo {author}
  {\bibfnamefont {S.}~\bibnamefont {Dubonos}}, \ and\ \bibinfo {author}
  {\bibfnamefont {A.}~\bibnamefont {Firsov}},\ }\href@noop {} {\bibfield
  {journal} {\bibinfo  {journal} {Nature}\ }\textbf {\bibinfo {volume} {438}},\
  \bibinfo {pages} {197} (\bibinfo {year} {2005})}\BibitemShut {NoStop}%
\bibitem [{\citenamefont {Lee}\ \emph {et~al.}(2008)\citenamefont {Lee},
  \citenamefont {Wei}, \citenamefont {Kysar},\ and\ \citenamefont
  {Hone}}]{lee2008measurement}%
  \BibitemOpen
  \bibfield  {author} {\bibinfo {author} {\bibfnamefont {C.}~\bibnamefont
  {Lee}}, \bibinfo {author} {\bibfnamefont {X.}~\bibnamefont {Wei}}, \bibinfo
  {author} {\bibfnamefont {J.~W.}\ \bibnamefont {Kysar}}, \ and\ \bibinfo
  {author} {\bibfnamefont {J.}~\bibnamefont {Hone}},\ }\href@noop {} {\bibfield
   {journal} {\bibinfo  {journal} {Science}\ }\textbf {\bibinfo {volume}
  {321}},\ \bibinfo {pages} {385} (\bibinfo {year} {2008})}\BibitemShut
  {NoStop}%
\bibitem [{\citenamefont {Lee}\ \emph {et~al.}(2013)\citenamefont {Lee},
  \citenamefont {Cooper}, \citenamefont {An}, \citenamefont {Lee},
  \citenamefont {van~der Zande}, \citenamefont {Petrone}, \citenamefont
  {Hammerberg}, \citenamefont {Lee}, \citenamefont {Crawford},\ and\
  \citenamefont {Oliver}}]{lee2013high}%
  \BibitemOpen
  \bibfield  {author} {\bibinfo {author} {\bibfnamefont {G.-H.}\ \bibnamefont
  {Lee}}, \bibinfo {author} {\bibfnamefont {R.~C.}\ \bibnamefont {Cooper}},
  \bibinfo {author} {\bibfnamefont {S.~J.}\ \bibnamefont {An}}, \bibinfo
  {author} {\bibfnamefont {S.}~\bibnamefont {Lee}}, \bibinfo {author}
  {\bibfnamefont {A.}~\bibnamefont {van~der Zande}}, \bibinfo {author}
  {\bibfnamefont {N.}~\bibnamefont {Petrone}}, \bibinfo {author} {\bibfnamefont
  {A.~G.}\ \bibnamefont {Hammerberg}}, \bibinfo {author} {\bibfnamefont
  {C.}~\bibnamefont {Lee}}, \bibinfo {author} {\bibfnamefont {B.}~\bibnamefont
  {Crawford}}, \ and\ \bibinfo {author} {\bibfnamefont {W.}~\bibnamefont
  {Oliver}},\ }\href@noop {} {\bibfield  {journal} {\bibinfo  {journal}
  {Science}\ }\textbf {\bibinfo {volume} {340}},\ \bibinfo {pages} {1073}
  (\bibinfo {year} {2013})}\BibitemShut {NoStop}%
\bibitem [{\citenamefont {Bunch}\ \emph {et~al.}(2008)\citenamefont {Bunch},
  \citenamefont {Verbridge}, \citenamefont {Alden}, \citenamefont {van~der
  Zande}, \citenamefont {Parpia}, \citenamefont {Craighead},\ and\
  \citenamefont {McEuen}}]{bunch2008impermeable}%
  \BibitemOpen
  \bibfield  {author} {\bibinfo {author} {\bibfnamefont {J.~S.}\ \bibnamefont
  {Bunch}}, \bibinfo {author} {\bibfnamefont {S.~S.}\ \bibnamefont
  {Verbridge}}, \bibinfo {author} {\bibfnamefont {J.~S.}\ \bibnamefont
  {Alden}}, \bibinfo {author} {\bibfnamefont {A.~M.}\ \bibnamefont {van~der
  Zande}}, \bibinfo {author} {\bibfnamefont {J.~M.}\ \bibnamefont {Parpia}},
  \bibinfo {author} {\bibfnamefont {H.~G.}\ \bibnamefont {Craighead}}, \ and\
  \bibinfo {author} {\bibfnamefont {P.~L.}\ \bibnamefont {McEuen}},\
  }\href@noop {} {\bibfield  {journal} {\bibinfo  {journal} {Nano letters}\
  }\textbf {\bibinfo {volume} {8}},\ \bibinfo {pages} {2458} (\bibinfo {year}
  {2008})}\BibitemShut {NoStop}%
\bibitem [{\citenamefont {Smith}\ \emph {et~al.}(2013)\citenamefont {Smith},
  \citenamefont {Niklaus}, \citenamefont {Paussa}, \citenamefont {Vaziri},
  \citenamefont {Fischer}, \citenamefont {Sterner}, \citenamefont {Forsberg},
  \citenamefont {Delin}, \citenamefont {Esseni},\ and\ \citenamefont
  {Palestri}}]{smith2013electromechanical}%
  \BibitemOpen
  \bibfield  {author} {\bibinfo {author} {\bibfnamefont {A.}~\bibnamefont
  {Smith}}, \bibinfo {author} {\bibfnamefont {F.}~\bibnamefont {Niklaus}},
  \bibinfo {author} {\bibfnamefont {A.}~\bibnamefont {Paussa}}, \bibinfo
  {author} {\bibfnamefont {S.}~\bibnamefont {Vaziri}}, \bibinfo {author}
  {\bibfnamefont {A.~C.}\ \bibnamefont {Fischer}}, \bibinfo {author}
  {\bibfnamefont {M.}~\bibnamefont {Sterner}}, \bibinfo {author} {\bibfnamefont
  {F.}~\bibnamefont {Forsberg}}, \bibinfo {author} {\bibfnamefont
  {A.}~\bibnamefont {Delin}}, \bibinfo {author} {\bibfnamefont
  {D.}~\bibnamefont {Esseni}}, \ and\ \bibinfo {author} {\bibfnamefont
  {P.}~\bibnamefont {Palestri}},\ }\href@noop {} {\bibfield  {journal}
  {\bibinfo  {journal} {Nano letters}\ }\textbf {\bibinfo {volume} {13}},\
  \bibinfo {pages} {3237} (\bibinfo {year} {2013})}\BibitemShut {NoStop}%
\bibitem [{\citenamefont {Lee}\ and\ \citenamefont
  {Feng}(2014)}]{lee2014atomically}%
  \BibitemOpen
  \bibfield  {author} {\bibinfo {author} {\bibfnamefont {J.}~\bibnamefont
  {Lee}}\ and\ \bibinfo {author} {\bibfnamefont {P.~X.-L.}\ \bibnamefont
  {Feng}},\ }in\ \href@noop {} {\emph {\bibinfo {booktitle} {Frequency Control
  Symposium (FCS), 2014 IEEE International}}}\ (\bibinfo {organization}
  {IEEE},\ \bibinfo {year} {2014})\ pp.\ \bibinfo {pages} {1--4}\BibitemShut
  {NoStop}%
\bibitem [{\citenamefont {Lee}\ \emph {et~al.}(2014)\citenamefont {Lee},
  \citenamefont {Wang}, \citenamefont {He}, \citenamefont {Shan},\ and\
  \citenamefont {Feng}}]{lee2014air}%
  \BibitemOpen
  \bibfield  {author} {\bibinfo {author} {\bibfnamefont {J.}~\bibnamefont
  {Lee}}, \bibinfo {author} {\bibfnamefont {Z.}~\bibnamefont {Wang}}, \bibinfo
  {author} {\bibfnamefont {K.}~\bibnamefont {He}}, \bibinfo {author}
  {\bibfnamefont {J.}~\bibnamefont {Shan}}, \ and\ \bibinfo {author}
  {\bibfnamefont {P.~X.-L.}\ \bibnamefont {Feng}},\ }\href@noop {} {\bibfield
  {journal} {\bibinfo  {journal} {Applied Physics Letters}\ }\textbf {\bibinfo
  {volume} {105}},\ \bibinfo {pages} {023104} (\bibinfo {year}
  {2014})}\BibitemShut {NoStop}%
\bibitem [{\citenamefont {Bao}\ and\ \citenamefont
  {Yang}(2007)}]{bao2007squeeze}%
  \BibitemOpen
  \bibfield  {author} {\bibinfo {author} {\bibfnamefont {M.}~\bibnamefont
  {Bao}}\ and\ \bibinfo {author} {\bibfnamefont {H.}~\bibnamefont {Yang}},\
  }\href@noop {} {\bibfield  {journal} {\bibinfo  {journal} {Sensors and
  Actuators A: Physical}\ }\textbf {\bibinfo {volume} {136}},\ \bibinfo {pages}
  {3} (\bibinfo {year} {2007})}\BibitemShut {NoStop}%
\bibitem [{\citenamefont {Andrews}\ \emph
  {et~al.}(1993{\natexlab{a}})\citenamefont {Andrews}, \citenamefont {Turner},
  \citenamefont {Harris},\ and\ \citenamefont {Harris}}]{andrews1993resonant}%
  \BibitemOpen
  \bibfield  {author} {\bibinfo {author} {\bibfnamefont {M.}~\bibnamefont
  {Andrews}}, \bibinfo {author} {\bibfnamefont {G.}~\bibnamefont {Turner}},
  \bibinfo {author} {\bibfnamefont {P.}~\bibnamefont {Harris}}, \ and\ \bibinfo
  {author} {\bibfnamefont {I.}~\bibnamefont {Harris}},\ }\href@noop {}
  {\bibfield  {journal} {\bibinfo  {journal} {Sensors and Actuators A:
  Physical}\ }\textbf {\bibinfo {volume} {36}},\ \bibinfo {pages} {219}
  (\bibinfo {year} {1993}{\natexlab{a}})}\BibitemShut {NoStop}%
\bibitem [{\citenamefont {Southworth}\ \emph {et~al.}(2009)\citenamefont
  {Southworth}, \citenamefont {Craighead},\ and\ \citenamefont
  {Parpia}}]{southworth2009pressure}%
  \BibitemOpen
  \bibfield  {author} {\bibinfo {author} {\bibfnamefont {D.}~\bibnamefont
  {Southworth}}, \bibinfo {author} {\bibfnamefont {H.}~\bibnamefont
  {Craighead}}, \ and\ \bibinfo {author} {\bibfnamefont {J.}~\bibnamefont
  {Parpia}},\ }\href@noop {} {\bibfield  {journal} {\bibinfo  {journal}
  {Applied Physics Letters}\ }\textbf {\bibinfo {volume} {94}},\ \bibinfo
  {pages} {213506} (\bibinfo {year} {2009})}\BibitemShut {NoStop}%
\bibitem [{\citenamefont {Andrews}\ \emph
  {et~al.}(1993{\natexlab{b}})\citenamefont {Andrews}, \citenamefont {Harris},\
  and\ \citenamefont {Turner}}]{andrews1993comparison}%
  \BibitemOpen
  \bibfield  {author} {\bibinfo {author} {\bibfnamefont {M.}~\bibnamefont
  {Andrews}}, \bibinfo {author} {\bibfnamefont {I.}~\bibnamefont {Harris}}, \
  and\ \bibinfo {author} {\bibfnamefont {G.}~\bibnamefont {Turner}},\
  }\href@noop {} {\bibfield  {journal} {\bibinfo  {journal} {Sensors and
  Actuators A: Physical}\ }\textbf {\bibinfo {volume} {36}},\ \bibinfo {pages}
  {79} (\bibinfo {year} {1993}{\natexlab{b}})}\BibitemShut {NoStop}%
\bibitem [{\citenamefont {Kumar}\ \emph {et~al.}(2015)\citenamefont {Kumar},
  \citenamefont {Reimann}, \citenamefont {Goossens}, \citenamefont {Besling},
  \citenamefont {Dolleman}, \citenamefont {Pijnenburg}, \citenamefont {van~der
  Avoort}, \citenamefont {Sarro},\ and\ \citenamefont
  {Steeneken}}]{kumar2015mems}%
  \BibitemOpen
  \bibfield  {author} {\bibinfo {author} {\bibfnamefont {L.}~\bibnamefont
  {Kumar}}, \bibinfo {author} {\bibfnamefont {K.}~\bibnamefont {Reimann}},
  \bibinfo {author} {\bibfnamefont {M.~J.}\ \bibnamefont {Goossens}}, \bibinfo
  {author} {\bibfnamefont {W.~F.}\ \bibnamefont {Besling}}, \bibinfo {author}
  {\bibfnamefont {R.~J.}\ \bibnamefont {Dolleman}}, \bibinfo {author}
  {\bibfnamefont {R.~H.}\ \bibnamefont {Pijnenburg}}, \bibinfo {author}
  {\bibfnamefont {C.}~\bibnamefont {van~der Avoort}}, \bibinfo {author}
  {\bibfnamefont {L.~P.}\ \bibnamefont {Sarro}}, \ and\ \bibinfo {author}
  {\bibfnamefont {P.~G.}\ \bibnamefont {Steeneken}},\ }\href@noop {} {\bibfield
   {journal} {\bibinfo  {journal} {Journal of Micromechanics and
  Microengineering}\ }\textbf {\bibinfo {volume} {25}},\ \bibinfo {pages}
  {045011} (\bibinfo {year} {2015})}\BibitemShut {NoStop}%
\bibitem [{\citenamefont {Castellanos-Gomez}\ \emph {et~al.}(2011)\citenamefont
  {Castellanos-Gomez}, \citenamefont {Wojtaszek}, \citenamefont {Tombros},
  \citenamefont {Agra{\"\i}t}, \citenamefont {van Wees},\ and\ \citenamefont
  {Rubio-Bollinger}}]{castellanos2011atomically}%
  \BibitemOpen
  \bibfield  {author} {\bibinfo {author} {\bibfnamefont {A.}~\bibnamefont
  {Castellanos-Gomez}}, \bibinfo {author} {\bibfnamefont {M.}~\bibnamefont
  {Wojtaszek}}, \bibinfo {author} {\bibfnamefont {N.}~\bibnamefont {Tombros}},
  \bibinfo {author} {\bibfnamefont {N.}~\bibnamefont {Agra{\"\i}t}}, \bibinfo
  {author} {\bibfnamefont {B.~J.}\ \bibnamefont {van Wees}}, \ and\ \bibinfo
  {author} {\bibfnamefont {G.}~\bibnamefont {Rubio-Bollinger}},\ }\href@noop {}
  {\bibfield  {journal} {\bibinfo  {journal} {Small}\ }\textbf {\bibinfo
  {volume} {7}},\ \bibinfo {pages} {2491} (\bibinfo {year} {2011})}\BibitemShut
  {NoStop}%
\bibitem [{\citenamefont {Castellanos-Gomez}\ \emph {et~al.}(2013)\citenamefont
  {Castellanos-Gomez}, \citenamefont {van Leeuwen}, \citenamefont {Buscema},
  \citenamefont {van~der Zant}, \citenamefont {Steele},\ and\ \citenamefont
  {Venstra}}]{castellanos2013single}%
  \BibitemOpen
  \bibfield  {author} {\bibinfo {author} {\bibfnamefont {A.}~\bibnamefont
  {Castellanos-Gomez}}, \bibinfo {author} {\bibfnamefont {R.}~\bibnamefont {van
  Leeuwen}}, \bibinfo {author} {\bibfnamefont {M.}~\bibnamefont {Buscema}},
  \bibinfo {author} {\bibfnamefont {H.~S.}\ \bibnamefont {van~der Zant}},
  \bibinfo {author} {\bibfnamefont {G.~A.}\ \bibnamefont {Steele}}, \ and\
  \bibinfo {author} {\bibfnamefont {W.~J.}\ \bibnamefont {Venstra}},\
  }\href@noop {} {\bibfield  {journal} {\bibinfo  {journal} {Advanced
  Materials}\ }\textbf {\bibinfo {volume} {25}},\ \bibinfo {pages} {6719}
  (\bibinfo {year} {2013})}\BibitemShut {NoStop}%
\bibitem [{\citenamefont {Cartamil-Bueno}\ \emph {et~al.}(2015)\citenamefont
  {Cartamil-Bueno}, \citenamefont {Steeneken}, \citenamefont {Tichelaar},
  \citenamefont {Navarro-Moratalla}, \citenamefont {Venstra}, \citenamefont
  {van Leeuwen}, \citenamefont {Coronado}, \citenamefont {van~der Zant},
  \citenamefont {Steele},\ and\ \citenamefont
  {Castellanos-Gomez}}]{cartamil2015high}%
  \BibitemOpen
  \bibfield  {author} {\bibinfo {author} {\bibfnamefont {S.~J.}\ \bibnamefont
  {Cartamil-Bueno}}, \bibinfo {author} {\bibfnamefont {P.~G.}\ \bibnamefont
  {Steeneken}}, \bibinfo {author} {\bibfnamefont {F.~D.}\ \bibnamefont
  {Tichelaar}}, \bibinfo {author} {\bibfnamefont {E.}~\bibnamefont
  {Navarro-Moratalla}}, \bibinfo {author} {\bibfnamefont {W.~J.}\ \bibnamefont
  {Venstra}}, \bibinfo {author} {\bibfnamefont {R.}~\bibnamefont {van
  Leeuwen}}, \bibinfo {author} {\bibfnamefont {E.}~\bibnamefont {Coronado}},
  \bibinfo {author} {\bibfnamefont {H.~S.}\ \bibnamefont {van~der Zant}},
  \bibinfo {author} {\bibfnamefont {G.~A.}\ \bibnamefont {Steele}}, \ and\
  \bibinfo {author} {\bibfnamefont {A.}~\bibnamefont {Castellanos-Gomez}},\
  }\href@noop {} {\bibfield  {journal} {\bibinfo  {journal} {Nano Research}\
  }\textbf {\bibinfo {volume} {8}},\ \bibinfo {pages} {2842} (\bibinfo {year}
  {2015})}\BibitemShut {NoStop}%
\bibitem [{\citenamefont {Bao}\ \emph {et~al.}(2002)\citenamefont {Bao},
  \citenamefont {Yang}, \citenamefont {Yin},\ and\ \citenamefont
  {Sun}}]{bao2002energy}%
  \BibitemOpen
  \bibfield  {author} {\bibinfo {author} {\bibfnamefont {M.}~\bibnamefont
  {Bao}}, \bibinfo {author} {\bibfnamefont {H.}~\bibnamefont {Yang}}, \bibinfo
  {author} {\bibfnamefont {H.}~\bibnamefont {Yin}}, \ and\ \bibinfo {author}
  {\bibfnamefont {Y.}~\bibnamefont {Sun}},\ }\href@noop {} {\bibfield
  {journal} {\bibinfo  {journal} {Journal of Micromechanics and
  Microengineering}\ }\textbf {\bibinfo {volume} {12}},\ \bibinfo {pages} {341}
  (\bibinfo {year} {2002})}\BibitemShut {NoStop}%
\bibitem [{\citenamefont {Rao}(2007)}]{rao2007vibration}%
  \BibitemOpen
  \bibfield  {author} {\bibinfo {author} {\bibfnamefont {S.~S.}\ \bibnamefont
  {Rao}},\ }\href@noop {} {\emph {\bibinfo {title} {Vibration of continuous
  systems}}}\ (\bibinfo  {publisher} {John Wiley \& Sons},\ \bibinfo {year}
  {2007})\BibitemShut {NoStop}%
\bibitem [{\citenamefont {Wah}(1962)}]{wah1962vibration}%
  \BibitemOpen
  \bibfield  {author} {\bibinfo {author} {\bibfnamefont {T.}~\bibnamefont
  {Wah}},\ }\href@noop {} {\bibfield  {journal} {\bibinfo  {journal} {the
  Journal of the Acoustical Society of America}\ }\textbf {\bibinfo {volume}
  {34}},\ \bibinfo {pages} {275} (\bibinfo {year} {1962})}\BibitemShut
  {NoStop}%
\bibitem [{\citenamefont {Veijola}\ \emph {et~al.}(1995)\citenamefont
  {Veijola}, \citenamefont {Kuisma}, \citenamefont {Lahdenper{\"a}},\ and\
  \citenamefont {Ryh{\"a}nen}}]{veijola1995equivalent}%
  \BibitemOpen
  \bibfield  {author} {\bibinfo {author} {\bibfnamefont {T.}~\bibnamefont
  {Veijola}}, \bibinfo {author} {\bibfnamefont {H.}~\bibnamefont {Kuisma}},
  \bibinfo {author} {\bibfnamefont {J.}~\bibnamefont {Lahdenper{\"a}}}, \ and\
  \bibinfo {author} {\bibfnamefont {T.}~\bibnamefont {Ryh{\"a}nen}},\
  }\href@noop {} {\bibfield  {journal} {\bibinfo  {journal} {Sensors and
  Actuators A: Physical}\ }\textbf {\bibinfo {volume} {48}},\ \bibinfo {pages}
  {239} (\bibinfo {year} {1995})}\BibitemShut {NoStop}%
\end{thebibliography}%


\providecommand*\mcitethebibliography{\thebibliography}
\csname @ifundefined\endcsname{endmcitethebibliography}
  {\let\endmcitethebibliography\endthebibliography}{}
\begin{mcitethebibliography}{8}
\providecommand*\natexlab[1]{#1}
\providecommand*\mciteSetBstSublistMode[1]{}
\providecommand*\mciteSetBstMaxWidthForm[2]{}
\providecommand*\mciteBstWouldAddEndPuncttrue
  {\def\EndOfBibitem{\unskip.}}
\providecommand*\mciteBstWouldAddEndPunctfalse
  {\let\EndOfBibitem\relax}
\providecommand*\mciteSetBstMidEndSepPunct[3]{}
\providecommand*\mciteSetBstSublistLabelBeginEnd[3]{}
\providecommand*\EndOfBibitem{}
\mciteSetBstSublistMode{f}
\mciteSetBstMaxWidthForm{subitem}{(\alph{mcitesubitemcount})}
\mciteSetBstSublistLabelBeginEnd
  {\mcitemaxwidthsubitemform\space}
  {\relax}
  {\relax}

\bibitem[Bunch et~al.(2008)Bunch, Verbridge, Alden, van~der Zande, Parpia,
  Craighead, and McEuen]{bunch2008impermeable}
Bunch,~J.~S.; Verbridge,~S.~S.; Alden,~J.~S.; van~der Zande,~A.~M.;
  Parpia,~J.~M.; Craighead,~H.~G.; McEuen,~P.~L. \emph{Nano letters}
  \textbf{2008}, \emph{8}, 2458--2462\relax
\mciteBstWouldAddEndPuncttrue
\mciteSetBstMidEndSepPunct{\mcitedefaultmidpunct}
{\mcitedefaultendpunct}{\mcitedefaultseppunct}\relax
\EndOfBibitem
\bibitem[Andrews et~al.(1993)Andrews, Harris, and
  Turner]{andrews1993comparison}
Andrews,~M.; Harris,~I.; Turner,~G. \emph{Sensors and Actuators A: Physical}
  \textbf{1993}, \emph{36}, 79--87\relax
\mciteBstWouldAddEndPuncttrue
\mciteSetBstMidEndSepPunct{\mcitedefaultmidpunct}
{\mcitedefaultendpunct}{\mcitedefaultseppunct}\relax
\EndOfBibitem
\bibitem[Bao and Yang(2007)Bao, and Yang]{bao2007squeeze}
Bao,~M.; Yang,~H. \emph{Sensors and Actuators A: Physical} \textbf{2007},
  \emph{136}, 3--27\relax
\mciteBstWouldAddEndPuncttrue
\mciteSetBstMidEndSepPunct{\mcitedefaultmidpunct}
{\mcitedefaultendpunct}{\mcitedefaultseppunct}\relax
\EndOfBibitem
\bibitem[Rao(2007)]{rao2007vibration}
Rao,~S.~S. \emph{Vibration of continuous systems}; John Wiley \& Sons,
  2007\relax
\mciteBstWouldAddEndPuncttrue
\mciteSetBstMidEndSepPunct{\mcitedefaultmidpunct}
{\mcitedefaultendpunct}{\mcitedefaultseppunct}\relax
\EndOfBibitem
\bibitem[Andrews et~al.(1993)Andrews, Turner, Harris, and
  Harris]{andrews1993resonant}
Andrews,~M.; Turner,~G.; Harris,~P.; Harris,~I. \emph{Sensors and Actuators A:
  Physical} \textbf{1993}, \emph{36}, 219--226\relax
\mciteBstWouldAddEndPuncttrue
\mciteSetBstMidEndSepPunct{\mcitedefaultmidpunct}
{\mcitedefaultendpunct}{\mcitedefaultseppunct}\relax
\EndOfBibitem
\bibitem[Wah(1962)]{wah1962vibration}
Wah,~T. \emph{the Journal of the Acoustical Society of America} \textbf{1962},
  \emph{34}, 275--281\relax
\mciteBstWouldAddEndPuncttrue
\mciteSetBstMidEndSepPunct{\mcitedefaultmidpunct}
{\mcitedefaultendpunct}{\mcitedefaultseppunct}\relax
\EndOfBibitem
\bibitem[Veijola et~al.(1995)Veijola, Kuisma, Lahdenper{\"a}, and
  Ryh{\"a}nen]{veijola1995equivalent}
Veijola,~T.; Kuisma,~H.; Lahdenper{\"a},~J.; Ryh{\"a}nen,~T. \emph{Sensors and
  Actuators A: Physical} \textbf{1995}, \emph{48}, 239--248\relax
\mciteBstWouldAddEndPuncttrue
\mciteSetBstMidEndSepPunct{\mcitedefaultmidpunct}
{\mcitedefaultendpunct}{\mcitedefaultseppunct}\relax
\EndOfBibitem
\end{mcitethebibliography}


\providecommand*\mcitethebibliography{\thebibliography}
\csname @ifundefined\endcsname{endmcitethebibliography}
  {\let\endmcitethebibliography\endthebibliography}{}
\begin{mcitethebibliography}{15}
\providecommand*\natexlab[1]{#1}
\providecommand*\mciteSetBstSublistMode[1]{}
\providecommand*\mciteSetBstMaxWidthForm[2]{}
\providecommand*\mciteBstWouldAddEndPuncttrue
  {\def\EndOfBibitem{\unskip.}}
\providecommand*\mciteBstWouldAddEndPunctfalse
  {\let\EndOfBibitem\relax}
\providecommand*\mciteSetBstMidEndSepPunct[3]{}
\providecommand*\mciteSetBstSublistLabelBeginEnd[3]{}
\providecommand*\EndOfBibitem{}
\mciteSetBstSublistMode{f}
\mciteSetBstMaxWidthForm{subitem}{(\alph{mcitesubitemcount})}
\mciteSetBstSublistLabelBeginEnd
  {\mcitemaxwidthsubitemform\space}
  {\relax}
  {\relax}

\bibitem[Novoselov et~al.(2005)Novoselov, Geim, Morozov, Jiang, Katsnelson,
  Grigorieva, Dubonos, and Firsov]{novoselov2005two}
Novoselov,~K.; Geim,~A.~K.; Morozov,~S.; Jiang,~D.; Katsnelson,~M.;
  Grigorieva,~I.; Dubonos,~S.; Firsov,~A. \emph{nature} \textbf{2005},
  \emph{438}, 197--200\relax
\mciteBstWouldAddEndPuncttrue
\mciteSetBstMidEndSepPunct{\mcitedefaultmidpunct}
{\mcitedefaultendpunct}{\mcitedefaultseppunct}\relax
\EndOfBibitem
\bibitem[Bolotin et~al.(2008)Bolotin, Sikes, Jiang, Klima, Fudenberg, Hone,
  Kim, and Stormer]{bolotin2008ultrahigh}
Bolotin,~K.~I.; Sikes,~K.; Jiang,~Z.; Klima,~M.; Fudenberg,~G.; Hone,~J.;
  Kim,~P.; Stormer,~H. \emph{Solid State Communications} \textbf{2008},
  \emph{146}, 351--355\relax
\mciteBstWouldAddEndPuncttrue
\mciteSetBstMidEndSepPunct{\mcitedefaultmidpunct}
{\mcitedefaultendpunct}{\mcitedefaultseppunct}\relax
\EndOfBibitem
\bibitem[Morozov et~al.(2008)Morozov, Novoselov, Katsnelson, Schedin, Elias,
  Jaszczak, and Geim]{morozov2008giant}
Morozov,~S.; Novoselov,~K.; Katsnelson,~M.; Schedin,~F.; Elias,~D.;
  Jaszczak,~J.; Geim,~A. \emph{Physical review letters} \textbf{2008},
  \emph{100}, 016602\relax
\mciteBstWouldAddEndPuncttrue
\mciteSetBstMidEndSepPunct{\mcitedefaultmidpunct}
{\mcitedefaultendpunct}{\mcitedefaultseppunct}\relax
\EndOfBibitem
\bibitem[Lee et~al.(2008)Lee, Wei, Kysar, and Hone]{lee2008measurement}
Lee,~C.; Wei,~X.; Kysar,~J.~W.; Hone,~J. \emph{science} \textbf{2008},
  \emph{321}, 385--388\relax
\mciteBstWouldAddEndPuncttrue
\mciteSetBstMidEndSepPunct{\mcitedefaultmidpunct}
{\mcitedefaultendpunct}{\mcitedefaultseppunct}\relax
\EndOfBibitem
\bibitem[Lee et~al.(2013)Lee, Cooper, An, Lee, van~der Zande, Petrone,
  Hammerberg, Lee, Crawford, and Oliver]{lee2013high}
Lee,~G.-H.; Cooper,~R.~C.; An,~S.~J.; Lee,~S.; van~der Zande,~A.; Petrone,~N.;
  Hammerberg,~A.~G.; Lee,~C.; Crawford,~B.; Oliver,~W. \emph{Science}
  \textbf{2013}, \emph{340}, 1073--1076\relax
\mciteBstWouldAddEndPuncttrue
\mciteSetBstMidEndSepPunct{\mcitedefaultmidpunct}
{\mcitedefaultendpunct}{\mcitedefaultseppunct}\relax
\EndOfBibitem
\bibitem[Bunch et~al.(2008)Bunch, Verbridge, Alden, van~der Zande, Parpia,
  Craighead, and McEuen]{bunch2008impermeable}
Bunch,~J.~S.; Verbridge,~S.~S.; Alden,~J.~S.; van~der Zande,~A.~M.;
  Parpia,~J.~M.; Craighead,~H.~G.; McEuen,~P.~L. \emph{Nano letters}
  \textbf{2008}, \emph{8}, 2458--2462\relax
\mciteBstWouldAddEndPuncttrue
\mciteSetBstMidEndSepPunct{\mcitedefaultmidpunct}
{\mcitedefaultendpunct}{\mcitedefaultseppunct}\relax
\EndOfBibitem
\bibitem[Smith et~al.(2013)Smith, Niklaus, Paussa, Vaziri, Fischer, Sterner,
  Forsberg, Delin, Esseni, and Palestri]{smith2013electromechanical}
Smith,~A.; Niklaus,~F.; Paussa,~A.; Vaziri,~S.; Fischer,~A.~C.; Sterner,~M.;
  Forsberg,~F.; Delin,~A.; Esseni,~D.; Palestri,~P. \emph{Nano letters}
  \textbf{2013}, \emph{13}, 3237--3242\relax
\mciteBstWouldAddEndPuncttrue
\mciteSetBstMidEndSepPunct{\mcitedefaultmidpunct}
{\mcitedefaultendpunct}{\mcitedefaultseppunct}\relax
\EndOfBibitem
\bibitem[Koenig et~al.(2011)Koenig, Boddeti, Dunn, and
  Bunch]{koenig2011ultrastrong}
Koenig,~S.~P.; Boddeti,~N.~G.; Dunn,~M.~L.; Bunch,~J.~S. \emph{Nature
  nanotechnology} \textbf{2011}, \emph{6}, 543--546\relax
\mciteBstWouldAddEndPuncttrue
\mciteSetBstMidEndSepPunct{\mcitedefaultmidpunct}
{\mcitedefaultendpunct}{\mcitedefaultseppunct}\relax
\EndOfBibitem
\bibitem[Andrews et~al.(1993)Andrews, Turner, Harris, and
  Harris]{andrews1993resonant}
Andrews,~M.; Turner,~G.; Harris,~P.; Harris,~I. \emph{Sensors and Actuators A:
  Physical} \textbf{1993}, \emph{36}, 219--226\relax
\mciteBstWouldAddEndPuncttrue
\mciteSetBstMidEndSepPunct{\mcitedefaultmidpunct}
{\mcitedefaultendpunct}{\mcitedefaultseppunct}\relax
\EndOfBibitem
\bibitem[Andrews et~al.(1993)Andrews, Harris, and
  Turner]{andrews1993comparison}
Andrews,~M.; Harris,~I.; Turner,~G. \emph{Sensors and Actuators A: Physical}
  \textbf{1993}, \emph{36}, 79--87\relax
\mciteBstWouldAddEndPuncttrue
\mciteSetBstMidEndSepPunct{\mcitedefaultmidpunct}
{\mcitedefaultendpunct}{\mcitedefaultseppunct}\relax
\EndOfBibitem
\bibitem[Kumar et~al.(2015)Kumar, Reimann, Goossens, Besling, Dolleman,
  Pijnenburg, van~der Avoort, Sarro, and Steeneken]{kumar2015mems}
Kumar,~L.; Reimann,~K.; Goossens,~M.~J.; Besling,~W.~F.; Dolleman,~R.~J.;
  Pijnenburg,~R.~H.; van~der Avoort,~C.; Sarro,~L.~P.; Steeneken,~P.~G.
  \emph{Journal of Micromechanics and Microengineering} \textbf{2015},
  \emph{25}, 045011\relax
\mciteBstWouldAddEndPuncttrue
\mciteSetBstMidEndSepPunct{\mcitedefaultmidpunct}
{\mcitedefaultendpunct}{\mcitedefaultseppunct}\relax
\EndOfBibitem
\bibitem[Castellanos-Gomez et~al.(2011)Castellanos-Gomez, Wojtaszek, Tombros,
  Agra{\"\i}t, van Wees, and Rubio-Bollinger]{castellanos2011atomically}
Castellanos-Gomez,~A.; Wojtaszek,~M.; Tombros,~N.; Agra{\"\i}t,~N.; van
  Wees,~B.~J.; Rubio-Bollinger,~G. \emph{Small} \textbf{2011}, \emph{7},
  2491--2497\relax
\mciteBstWouldAddEndPuncttrue
\mciteSetBstMidEndSepPunct{\mcitedefaultmidpunct}
{\mcitedefaultendpunct}{\mcitedefaultseppunct}\relax
\EndOfBibitem
\bibitem[Castellanos-Gomez et~al.(2013)Castellanos-Gomez, van Leeuwen, Buscema,
  van~der Zant, Steele, and Venstra]{castellanos2013single}
Castellanos-Gomez,~A.; van Leeuwen,~R.; Buscema,~M.; van~der Zant,~H.~S.;
  Steele,~G.~A.; Venstra,~W.~J. \emph{Advanced Materials} \textbf{2013},
  \emph{25}, 6719--6723\relax
\mciteBstWouldAddEndPuncttrue
\mciteSetBstMidEndSepPunct{\mcitedefaultmidpunct}
{\mcitedefaultendpunct}{\mcitedefaultseppunct}\relax
\EndOfBibitem
\bibitem[Cartamil-Bueno et~al.(2015)Cartamil-Bueno, Steeneken, Tichelaar,
  Navarro-Moratalla, Venstra, van Leeuwen, Coronado, van~der Zant, Steele, and
  Castellanos-Gomez]{cartamil2015high}
Cartamil-Bueno,~S.~J.; Steeneken,~P.~G.; Tichelaar,~F.~D.;
  Navarro-Moratalla,~E.; Venstra,~W.~J.; van Leeuwen,~R.; Coronado,~E.; van~der
  Zant,~H.~S.; Steele,~G.~A.; Castellanos-Gomez,~A. \emph{arXiv preprint
  arXiv:1505.04727} \textbf{2015}, \relax
\mciteBstWouldAddEndPunctfalse
\mciteSetBstMidEndSepPunct{\mcitedefaultmidpunct}
{}{\mcitedefaultseppunct}\relax
\EndOfBibitem
\end{mcitethebibliography}

\end{document}